\documentclass[12pt]{amsart}
\usepackage[dvipsnames]{xcolor}
\definecolor{darkblue}{rgb}{0.0, 0.0, 0.61}
\usepackage[colorlinks=true,linkcolor=black,citecolor=darkblue,urlcolor=darkblue]{hyperref}

\usepackage{amsfonts,amsmath,amsthm,amssymb}
\usepackage[mathscr]{eucal}
\usepackage{tikz}
\usetikzlibrary{arrows,snakes}
\usepackage{algorithm}
\usepackage[noend]{algpseudocode}
\usepackage[normalem]{ulem}

\usepackage{amsmath,amsthm}
\newtheorem{thm}{Theorem}[section]

\newtheorem{clm}[thm]{Claim}
\theoremstyle{definition}
\newtheorem{df}[thm]{Definition}

\theoremstyle{remark}

\numberwithin{equation}{section}
\newcommand{\thref}[1]{Theorem~\ref{#1}}
\newcommand{\secref}[1]{\S\ref{#1}}
\newcommand{\lmref}[1]{Lemma~\ref{#1}}
\DeclareMathOperator{\satf}{\models}
\DeclareMathOperator{\miff}{\Longleftrightarrow} 
\DeclareMathOperator{\mimplies}{\Longrightarrow} 
\DeclareMathOperator{\mimplied}{\Longleftarrow}  
\newcommand{\set}[2]{\{#1\,:\,\text{#2}\}} 
\DeclareMathOperator{\id}{\textsl{id}}     
\DeclareMathOperator{\miffdef}{\stackrel{def}{\miff}}
\DeclareMathOperator{\eqdef}{\stackrel{def}{=}}

\DeclareMathOperator{\neareq}{\stackrel{\ae}{=}}
\DeclareMathOperator{\aeqv}{\stackrel{ae}{\equiv}}
\DeclareMathOperator{\aeq}{\stackrel{ae}{=}}
\DeclareMathOperator{\BZ}{\mathbb{Z}} 
\DeclareMathOperator{\BQ}{\mathbb{Q}} 
\DeclareMathOperator{\BR}{\mathbb{R}} 
\DeclareMathOperator{\BC}{\mathbb{C}} 
\DeclareMathOperator{\re}{Re}         
\DeclareMathOperator{\im}{Im}         
\DeclareMathOperator{\Ker}{Ker}
\DeclareMathOperator{\meet}{\wedge}
\DeclareMathOperator{\join}{\vee}
\DeclareMathOperator{\union}{\cup}
\DeclareMathOperator{\inter}{\cap}  
\DeclareMathOperator{\iso}{\cong}   
\DeclareMathOperator{\piso}{\simeq} 
\newcommand{\Star}{^ *}
\DeclareMathOperator{\compose}{{\scriptstyle \circ}} 
\newcommand{\wec}[1]{{\mathbf{#1}}}  
\newcommand{\wrel}[1]{\;#1\;}     
\newcommand{\m}[1]{{\mathbf{\uppercase{#1}}}}
\newcommand{\vr}[1]{{\mathcal{\uppercase {#1}}}}
\DeclareMathOperator{\Var}{\mathbf{V}}
\DeclareMathOperator{\Ho}{\mathbf{H}}
\DeclareMathOperator{\Su}{\mathbf{S}}
\DeclareMathOperator{\Pd}{\mathbf{P}}
\DeclareMathOperator{\Iso}{\mathbf{I}}
\DeclareMathOperator{\Ps}{\mathbf{P}_s}
\newcommand{\fin}[1]{{\mathbf{#1}}_{\text{\upshape fin}}}
\DeclareMathOperator{\Pol}{Pol}
\DeclareMathOperator{\Clo}{Clo}
\DeclareMathOperator{\Con}{Con}
\DeclareMathOperator{\Conb}{\mathbf{Con}} 
\DeclareMathOperator{\cg}{Cg}
\DeclareMathOperator{\sg}{Sg}
\newcommand{\Cg}[1]{{\cg}^{\m{#1}}}
\newcommand{\Sg}[1]{{\sg}^{\m{#1}}}
\newcommand{\M}[3]{\text{\upshape M}_{\m{#1}} (#2, #3)}
\newcommand{\minl}[1]{\text{\upshape Min}_{\m{#1}}}
\DeclareMathOperator{\typ}{typ}
\DeclareMathOperator{\ann}{ann}
\DeclareMathOperator{\Sep}{Sep}
\DeclareMathOperator{\ssb}{ssb}
\DeclareMathOperator{\sst}{sst}
\DeclareMathOperator{\utyp}{\mathbf{1}}
\DeclareMathOperator{\atyp}{\mathbf{2}}
\DeclareMathOperator{\btyp}{\mathbf{3}}
\DeclareMathOperator{\ltyp}{\mathbf{4}}
\DeclareMathOperator{\styp}{\mathbf{5}}
\DeclareMathOperator{\ityp}{\mathbf{i}}
\DeclareMathOperator{\jtyp}{\mathbf{j}}
\DeclareMathOperator{\solv}{\stackrel{s}{\sim}}
\DeclareMathOperator{\ssolv}{\stackrel{ss}{\sim}}
\newcommand{\bd}{\begin{description}}
\newcommand{\ed}{\end{description}}
\newcommand{\lb}{\langle} 
\newcommand{\rb}{\rangle}
\newcommand{\lv}{{[\![}}  
\newcommand{\rv}{{]\!]}}
\newcommand{\ul}{\underline}
\newcommand{\ol}{\overline}
\newcommand{\vep}{\varepsilon}
\newcommand{\di}{\diamond}
\newcommand{\alert}[1]{{\color{red} #1}}

\newcommand{\tup}[1]{\mathbf{#1}}
\newcommand{\ra}{\rightarrow}
\newcommand{\Ra}{\Rightarrow}
\newcommand{\ran}{\mathrm{ran}}
\newcommand{\dom}{\mathrm{dom}}
\renewcommand{\algorithmicrequire}{\textbf{Input:}}
\newcommand{\barf}{\smash{\overline{f}}}
\newcommand{\Change}{\mbox{\textsc{Change}}}
\newcommand{\ind}{\rule{.25in}{0in}}
\newcommand{\Sink}{\mbox{\textsc{Sink}}}
\newcommand{\RemoveNotMinority}{\mbox{\textsc{RemoveNotMinority}}}
\newcommand{\RecursiveRNM}{\mbox{\textsc{RecursiveRNM}}}
\newcommand{\SmallInstance}{\mbox{\textsc{Small-Instance}}}
\newcommand{\BigInstance}{\mbox{\textsc{Big-Instance}}}
\newcommand{\UpdateF}{\mbox{\textsc{Update-f}}}
\newcommand{\RemoveA}{\mbox{\textsc{Remove-a}}}
\newcommand{\Stack}{\mbox{\textsc{Stack}}}
\newcommand{\false}{\mbox{\textsc{false}}}
\newcommand{\true}{\mbox{\textsc{true}}}
\newcommand{\Pop}{\mbox{\textsc{Pop}}}
\newcommand{\Visit}{\mathit{Visit}}

\newcommand{\Preprocessing}{\mbox{\textsc{PreProcessing}}}
\newcommand{\NoHom}{\mbox{\textsc{NoHom}}}
\newcommand{\Output}{\mbox{\textsc{Output}}}
\newcommand{\proj}{\mathrm{pr}}
\newcommand{\scrL}{\mathscr{L}}
\newcommand{\barG}{G}
\newcommand{\barGL}{\overline{G}_L}
\newcommand{\barGLone}{\overline{G}_{L_1}}
\newcommand{\barw}{\overline{w}}
\newcommand{\bbH}{\mathbb H}
\newcommand{\bbE}{\mathbb E}
\newcommand{\bbP}{\mathbb P}
\newcommand{\bbQ}{\mathbb Q}
\newcommand{\bbS}{\mathbb S}
\newcommand{\bbT}{{\mathbb S}'}
\newcommand{\bbG}{\mathbb G}
\newcommand{\bbA}{\mathbb A}
\newcommand{\bara}{\overline{a}}
\newcommand{\barb}{\overline{b}}
\newcommand{\barc}{\overline{c}}
\newcommand{\bard}{\overline{d}}
\newcommand{\bare}{\overline{e}}
\newcommand{\bart}{\overline{t}}
\newcommand{\bbZ}{\mathbb Z}
\newcommand{\CSP}{\mathrm{CSP}}
\newcommand{\Hom}{\mathrm{Hom}}

\newcommand{\varA}{a}  

\newcommand{\verta}{\alpha}
\newcommand{\vertb}{\beta}
\newcommand{\vertc}{\gamma}
\newcommand{\vertd}{\delta}
\newcommand{\vertt}{\tau}
\newcommand{\varR}{\lambda}   
\newcommand{\calI}{\mathscr{I}}
\newcommand{\barv}{\overline{v}}

\newcommand{\barverta}{\overline{\verta}}  
\newcommand{\barvertb}{\overline{\vertb}}
\newcommand{\barvertc}{\overline{\vertc}}
\newcommand{\barvertd}{\overline{\vertd}}
\newcommand{\barvertt}{\overline{\vertt}}
\newcommand{\barvarR}{\overline{\varR}}

\newcommand{\vertH}{u}
\newcommand{\vertG}{v}

\newcommand{\lab}[3]{\scriptstyle{\vertH_{#1 #2}^{#3}}}
\newcommand{\labL}[3]{\scriptstyle{\vertH_{#1 #2}^{#3 L}}}
\newcommand{\labR}[3]{\scriptstyle{\vertH_{#1 #2}^{#3 R}}}

\newcommand{\biglab}[3]{\vertH_{#1 #2}^{#3}}
\newcommand{\biglabL}[3]{\vertH_{#1 #2}^{#3 L}}
\newcommand{\biglabR}[3]{\vertH_{#1 #2}^{#3 R}}

\newcommand{\laba}[3]{\scriptstyle{\vertH_{#1 \alpha}^{#3}}}
\newcommand{\labLa}[3]{\scriptstyle{\vertH_{#1 \alpha}^{#3 L}}}
\newcommand{\labRa}[3]{\scriptstyle{\vertH_{#1 \alpha}^{#3 R}}}

\newcommand{\labb}[3]{\scriptstyle{\vertH_{#1 \beta}^{#3}}}
\newcommand{\labLb}[3]{\scriptstyle{\vertH_{#1 \beta}^{#3 L}}}
\newcommand{\labRb}[3]{\scriptstyle{\vertH_{#1 \beta}^{#3 R}}}

\newcommand{\labc}[3]{\scriptstyle{\vertH_{#1 \gamma}^{#3}}}
\newcommand{\labLc}[3]{\scriptstyle{\vertH_{#1 \gamma}^{#3 L}}}
\newcommand{\labRc}[3]{\scriptstyle{\vertH_{#1 \gamma}^{#3 R}}}

\newcommand{\labd}[3]{\scriptstyle{\vertH_{#1 \delta}^{#3}}}
\newcommand{\labLd}[3]{\scriptstyle{\vertH_{#1 \delta}^{#3 L}}}
\newcommand{\labRd}[3]{\scriptstyle{\vertH_{#1 \delta}^{#3 R}}}

\newcommand{\labt}[3]{\scriptstyle{\vertH_{#1 \tau}^{#3}}}
\newcommand{\labLt}[3]{\scriptstyle{\vertH_{#1 \tau}^{#3 L}}}
\newcommand{\labRt}[3]{\scriptstyle{\vertH_{#1 \tau}^{#3 R}}}

\newcommand{\labl}[3]{\scriptstyle{\vertH_{#1 \lambda}^{#3}}}
\newcommand{\labLl}[3]{\scriptstyle{\vertH_{#1 \lambda}^{#3 L}}}
\newcommand{\labRl}[3]{\scriptstyle{\vertH_{#1 \lambda}^{#3 R}}}

\newcommand{\qlab}[4]{\scriptstyle{\vertG_{#1}^{#4}}}
\newcommand{\qlabL}[4]{\scriptstyle{\vertG_{#1}^{#4 L}}}
\newcommand{\qlabR}[4]{\scriptstyle{\vertG_{#1}^{#4 R}}}

\newcommand{\bigqlab}[2]{\vertG_{#1}^{#2}}
\newcommand{\bigqlabL}[2]{\vertG_{#1}^{#2 L}}
\newcommand{\bigqlabR}[2]{\vertG_{#1}^{#2 R}}
\newcommand{\bigqlabS}[2]{\vertG_{#1}^{#2\ast}}

\newcommand{\bigqlabj}[3]{\vertG_{#1,#3}^{#2}}
\newcommand{\bigqlabLj}[3]{\vertG_{#1,#3}^{#2 L}}
\newcommand{\bigqlabRj}[3]{\vertG_{#1,#3}^{#2 R}}
\newcommand{\bigqlabSj}[3]{\vertG_{#1,#3}^{#2\ast}}

\newcommand{\Qtop}{t}
\newcommand{\Qbot}{\bot}

\newcommand{\bbQcopy}[2]{\bbQ_{#1,#2}}
\newcommand{\graph}{\mathrm{graph}}

\begin{document}

\title{Refuting Feder, Kinne and Rafiey}

\author{Ross Willard}
\date{July 28, 2017}
\thanks{The support of the Natural Sciences and Engineering Research Council
(NSERC) of Canada is gratefully acknowledged.}

\begin{abstract}
I give an 
example showing that the recent claimed solution by Feder, Kinne and Rafiey
to the CSP Dichotomy Conjecture is not correct.
\end{abstract}

\maketitle

\section{The Feder-Kinne-Rafiey algorithm}

In January 2017, Feder, Kinne and Rafiey posted a manuscript \cite{rkf}
on the arXiv 
which purportedly gives a polynomial-time
algorithm solving the digraph homomorphism problem $\Hom(\bbH)$,
assuming only that the target digraph $\bbH$
has a weak near unanimity (WNU) polymorphism.  
If correct, their algorithm would finish the proof of the 
CSP Dichotomy Conjecture of Feder and Vardi \cite{fv}.  

In my opinion, their manuscript was imprecise (or worse) at many points, 
making it difficult for readers to ascertain \emph{what} exactly was their algorithm.
The authors subsequently posted two new versions of their manuscript, 
the most
recent (v3) on July 1, 2017.  All three manuscripts attempt to articulate
implementations of
an algorithm-idea which roughly speaking is the following:
given an instance  $\bbG\stackrel{?}{\ra}\bbH$ for which the target digraph
$\bbH$ has the WNU polymorphism $\phi$:
\begin{enumerate}
\item
Do some local consistency checking. This will produce, for each vertex $x \in V(\bbG)$,
a list $L(x) \subseteq V(\bbH)$ from which ``provably impossible values for $h(x)$"
(here $h$ is a hypothetical homomorphism)
have been removed.  (To be precise: the algorithm enforces \emph{path-consistency},
a.k.a.\ (2,3)-consistency.  The lists $L(x)$ are the cells, a.k.a.\ ``potatoes,"
of the resulting \emph{microstructure graph}.)
\item
Associate to each vertex $x \in V(\bbG)$ the restriction $\phi_x:=\phi|_{L(x)}$
 to $L(x)$ of the WNU polymorphism.
\end{enumerate}
With this preprocessing out of the way, the main part of the algorithm-idea 
is to iteratively ``shrink" the
lists $L(x)$, always maintaining the property that
if solutions exist, then at least one solution will
live in $\prod_{x}L(x)$.  This will purportedly be accomplished by:
\begin{enumerate}
\setcounter{enumi}{2}
\item
Allowing the maps $\phi_x$ to evolve (in some prescribed way).
At all stages, each
$\phi_x$ must remain a not-necessarily-idempotent WNU operation on $L(x)$, and the
family $(\phi_x:x \in V(\bbG))$ must be a ``multi-sorted polymorphism" of the
microstructure graph defined by $(\bbG,\bbH,L)$.
\item
Using the evolved maps $(\phi_x:x \in V(\bbG))$ to 
predict $x \in V(\bbG)$ and $a \in L(x)$ such that $a$ can be safely deleted from 
the list $L(x)$.
\end{enumerate}

The novelty of the Feder-Kinne-Rafiey algorithm-idea is in steps (3) and 
(4).  The details of step (3) are delicate (and murky).
However, the authors' criterion in step (4) is very clear:
if the current list $L(x)$ has elements $a,b$ such that the current
$\phi_x$ satisfies $\phi_x(b,b,\ldots,b,a)\ne a$, then $a$
can be safely deleted from $L(x)$.  

If this algorithm-idea could be implemented, then it would reduce an instance
of $\Hom(\bbH)$
to testing consistency of a microstructure graph with a multi-sorted WNU polymorphism
$(\phi_x:x \in V(\bbG))$ where each $\phi_x$ satisfies $\phi_x(b,b,\ldots,b,a)=a$
for all $a,b \in L(x)$.  
This latter problem is known to CSPers as the ``multi-sorted Mal'tsev case" and 
there are known algorithms solving it
\cite{bul-mal,bd}.  Hence the Feder-Kinne-Rafiey algorithm-idea aims
to reduce the WNU case of the general digraph homomorphism problem 
to the multi-sorted Mal'tsev case.

Meanwhile, Bulatov \cite{bul-DC} and Zhuk \cite{zhuk} have independently announced solutions
to the CSP Dichotomy Conjecture.  Their manuscripts are much more complicated
than those of Feder, Kinne and Rafiey.  Thus it
is of interest whether the ideas of Feder, Kinne and Rafiey are correct (or at 
least correctable).
In this note I will show that,
unfortunately, their algorithm-idea cannot be implemented, because
their criterion for shrinking a list (step 4) cannot generally be done safely.

\begin{clm} \label{fkr-clm}
There exists a finite digraph $\bbH$ with a $3$-ary WNU polymorphism
$\phi'$, and another finite digraph $\bbG$, such that 
\begin{enumerate}
\item
There exists a unique homomorphism $h:\bbG\ra \bbH$.
\item
If $L(x)$ ($x \in V(\bbG)$) are the lists 
obtained by enforcing $(2,3)$-consistency for $\bbG\stackrel{?}{\ra}\bbH$, 
then for all $x \in V(\bbG)$,
\begin{enumerate}
\item
There exist $a,b \in L(x)$ such that
$\phi'(b,b,\ldots,b,a)\ne a$.  
\item
For all $a,b \in L(x)$,  if $\phi'(b,b,\ldots,b,a)\ne a$ then
$h(x)=a$.  
\end{enumerate}
\end{enumerate}
Moreover, $\bbH$ can be chosen to be core and balanced.
\end{clm}

Note that condition (a) says that, after preprocessing the instance
$\bbG\stackrel{?}{\ra}\bbH$, 
we are not yet in the Mal'tsev case (with respect to $\phi'$), while
condition (b) says that any attempt to apply the criterion from step (4)
of the Feder-Kinne-Rafiey algorithm-idea (with $\phi'$)
will lead to a smaller list that is disjoint from the unique solution.
In particular, this implies that
no matter how the details of step (3) are
implemented, the algorithm-idea articulated in any of versions 
(v1)--(v3) of the Feder-Kinne-Rafiey manuscript 
 will fail to give the correct answer on input
$\bbH,\bbG$ (using this $\phi'$).

The proof of Claim~\ref{fkr-clm} occupies the next section.
In the final section of this note, I will give further examples that illustrate
some challenges that any attempt to modify the algorithm-idea of Feder,
Kinne and Rafiey must necessarily overcome.

\section{Proof of Claim~\ref{fkr-clm}} 

\subsection{High-level description for CSP experts} \label{subsec-high}

The construction of $\bbH$, $\bbG$ and $\phi$ is based on a small CSP instance
over the 
template $\bbA=\langle A,R\rangle$ with domain
$A=\{0,1,2\}$ and a single 5-ary relation 
\[
R=\{(0,0,0,1,0),(0,1,1,0,0),(1,0,1,0,0), (1,1,0,1,0),(2,2,2,2,0)\}.
\]
$\bbA$ is core, in fact has no nontrivial endomorphisms.
Note that $R \cap \{0,1\}^5$ is an affine subspace of $(\bbZ_2)^5$ and
so is closed under $m(x,y,z) := 
x+y+z$ (mod 2) applied coordinatewise.  
Using this fact, it is not hard to see that the operation
$\phi:A^3\ra A$ is a polymorphism of $\bbA$:

\[
\phi(x,y,z) = \left\{\begin{array}{cl}
m(x,y,z) & \mbox{if $\{x,y,z\} \subseteq \{0,1\}$}\\
a & \mbox{if $(x,y,z) \in \{(2,a,b),(b,2,a),(a,b,2)\}$ with $a \in \{0,1\}$}\\
2 & \mbox{if $x=y=z=2$.}
\end{array}\right.
\]
Clearly $\phi(x,x,x)=x$ and $\phi(x,y,z) = \phi(y,z,x)$ for all 
$x,y,z \in A$, which means $\phi$ is idempotent and cyclic (and hence is a WNU).

Let $R_1$ and $R_2$ be the 3-ary relations obtained by projecting $R$ onto
coordinates $(1,2,3)$ and $(1,2,4)$ respectively,
and let $\bbA' = \langle A,R_1,R_2\rangle$.
The CSP($\bbA'$) instance $\calI$ given by
\[
R_1(x_1,x_2,x_3) \quad\&\quad R_1(x_1,x_5,x_6) \quad\&\quad
R_1(x_2,x_4,x_6) \quad\&\quad R_2(x_3,x_4,x_5)
\]
is (2,3)-minimal, meaning that the lists obtained after enforcing
(2,3)-consistency are $L(x_i)=A$ for all $i$.  Moreover, 
the constant map $\{x_1,\ldots,x_6\}\mapsto 2$ is the unique solution
to $\calI$, 
which means that $\calI$ has a solution, but has no solution satisfying
$h$ satisfying $h(x_i)\ne 2$.
Yet for all $a,b \in A$, $\phi(b,b,a)\ne a$ iff $a=2$ and $b \in \{0,1\}$.


Using a 
construction of Bul\'{i}n, Deli\'{c}, Jackson and Niven \cite{bdjn} (improving
Feder and Vardi \cite{fv}),
one can translate $\bbA$, $\calI$ and $\phi$
 to a digraph $\bbH$, an instance
$\bbG$ of $\Hom(\bbH)$, and a polymorphism $\phi'$ of $\bbH$
 with essentially the same properties.
I will give a detailed description of the resulting digraphs $\bbH,\bbG$
in sections~\ref{sec-H} and~\ref{sec-G}

\subsection{Additional remarks}
$\bbA$ and $\bbA'$  are examples of a kind of template that Bulatov \cite{bul-block} calls
\emph{semilattice block Mal'tsev}.  This refers to the fact
that the domain $\{0,1,2\}$ has
a partition, namely $01|2$, which is invariant under $\phi$ and is such that
$\phi$ is a Mal'tsev operation on each block; and $\bbA$ (or $\bbA'$) additionally
has a 2-ary polymorphism $f$ (namely, $f(x,y) := \phi(x,x,y)$) which respects
this partition and has the
property that, modulo the partition, $f$ is a semilattice operation.
Semilattice block Mal'tsev templates have been a sticking point for researchers
attempting to solve the CSP Dichotomy Conjecture.  In particular, 
Mar\'{o}ti has circulated but never published some important partial results
\cite{miklos},
and Markovi\'{c} and McKenzie \cite{mar-mck} and Payne
\cite{payne} have some further unpublished results.
It was only earlier this year that Bulatov \cite{bul-block}
finally solved the Dichotomy Conjecture for 
semilattice block Mal'tsev templates, two months
before he announced his solution to the full Dichotomy Conjecture.
So it is perhaps not a surprise that semilattice block Mal'tsev templates should
be the source of counter-examples to the algorithm-idea of Feder, Kinne and 
Rafiey.

\subsection{The digraph $\bbH$} \label{sec-H}

The following construction based on $\bbA$ is due to
Bul\'in, Deli\'c, Jackson and Niven \cite{bdjn}.
Define a balanced digraph $\bbH$ as follows.  The vertex set $V(\bbH)$ 
consists of 
the union of $\{0,1,2\}$, $\{\verta,\vertb,\vertc,\vertd,\vertt\}$, 
and a set of 190 auxiliary vertices lying
on 15 pairwise disjoint oriented paths of net length 7 connecting each
pair in
$\{0,1,2\}\times\{\verta,\vertb,\vertc,\vertd,\vertt\}$.  These paths are:

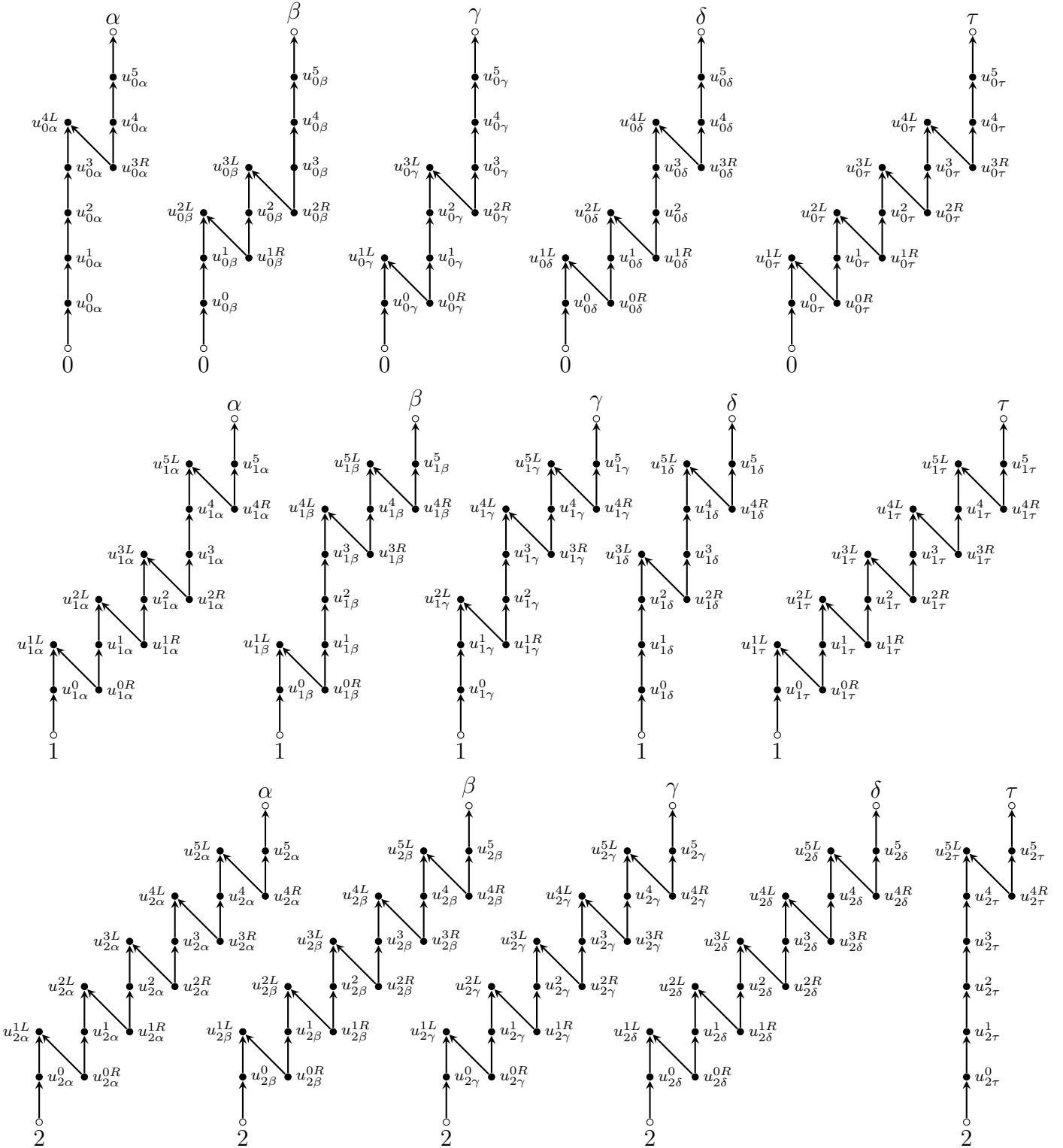
\hspace{-.6in}
\begin{tikzpicture}[scale=.8]

\draw  (0,0) circle (2.0pt);
\draw [fill] (0,1) circle (2.0pt);
\draw [fill] (0,2) circle (2.0pt);
\draw [fill] (0,3) circle (2.0pt);
\draw [fill] (0,4) circle (2.0pt);
\draw [fill] (0,5) circle (2.0pt);
\draw [fill] (1,4) circle (2.0pt);
\draw [fill] (1,5) circle (2.0pt);
\draw [fill] (1,6) circle (2.0pt);
\draw  (1,7) circle (2.0pt);
\draw [->,>=stealth,thick,shorten >=2pt,shorten <=2pt] (0,0) -- (0,1);
\draw [->,>=stealth,thick,shorten >=2pt,shorten <=2pt] (0,1) -- (0,2);
\draw [->,>=stealth,thick,shorten >=2pt,shorten <=2pt] (0,2) -- (0,3);
\draw [->,>=stealth,thick,shorten >=2pt,shorten <=2pt] (0,3) -- (0,4);
\draw [->,>=stealth,thick,shorten >=2pt,shorten <=2pt] (0,4) -- (0,5);
\draw [->,>=stealth,thick,shorten >=2pt,shorten <=2pt] (1,4) -- (0,5);
\draw [->,>=stealth,thick,shorten >=2pt,shorten <=2pt] (1,4) -- (1,5);
\draw [->,>=stealth,thick,shorten >=2pt,shorten <=2pt] (1,5) -- (1,6);
\draw [->,>=stealth,thick,shorten >=2pt,shorten <=2pt] (1,6) -- (1,7);
\node [anchor=north] at (0,0) {$0$};
\node [anchor=west] at (0,1) {$\laba 0a0$};
\node [anchor=west] at (0,2) {$\laba 0a1$};
\node [anchor=west] at (0,3) {$\laba 0a2$};
\node [anchor=west] at (0,4) {$\laba 0a3$};
\node [anchor=east] at (0,5) {$\labLa 0a4$};
\node [anchor=west] at (1,4) {$\labRa 0a3$};
\node [anchor=west] at (1,5) {$\laba 0a4$};
\node [anchor=west] at (1,6) {$\laba 0a5$};
\node [anchor=south] at (1,7) {$\verta$};

\draw  (3,0) circle (2.0pt);
\draw [fill] (3,1) circle (2.0pt);
\draw [fill] (3,2) circle (2.0pt);
\draw [fill] (3,3) circle (2.0pt);
\draw [fill] (4,2) circle (2.0pt);
\draw [fill] (4,3) circle (2.0pt);
\draw [fill] (4,4) circle (2.0pt);
\draw [fill] (5,3) circle (2.0pt);
\draw [fill] (5,4) circle (2.0pt);
\draw [fill] (5,5) circle (2.0pt);
\draw [fill] (5,6) circle (2.0pt);
\draw  (5,7) circle (2.0pt);
\draw [->,>=stealth,thick,shorten >=2pt,shorten <=2pt] (3,0) -- (3,1);
\draw [->,>=stealth,thick,shorten >=2pt,shorten <=2pt] (3,1) -- (3,2);
\draw [->,>=stealth,thick,shorten >=2pt,shorten <=2pt] (3,2) -- (3,3);
\draw [->,>=stealth,thick,shorten >=2pt,shorten <=2pt] (4,2) -- (3,3);
\draw [->,>=stealth,thick,shorten >=2pt,shorten <=2pt] (4,2) -- (4,3);
\draw [->,>=stealth,thick,shorten >=2pt,shorten <=2pt] (4,3) -- (4,4);
\draw [->,>=stealth,thick,shorten >=2pt,shorten <=2pt] (5,3) -- (4,4);
\draw [->,>=stealth,thick,shorten >=2pt,shorten <=2pt] (5,3) -- (5,5);
\draw [->,>=stealth,thick,shorten >=2pt,shorten <=2pt] (5,4) -- (5,5);
\draw [->,>=stealth,thick,shorten >=2pt,shorten <=2pt] (5,5) -- (5,6);
\draw [->,>=stealth,thick,shorten >=2pt,shorten <=2pt] (5,6) -- (5,7);
\node [anchor=north] at (3,0) {$0$};
\node [anchor=west] at (3,1) {$\labb 0b0$};
\node [anchor=west] at (3,2) {$\labb 0b1$};
\node [anchor=east] at (3,3) {$\labLb 0b2$};
\node [anchor=west] at (4,2) {$\labRb 0b1$};
\node [anchor=west] at (4,3) {$\labb 0b2$};
\node [anchor=east] at (4,4) {$\labLb 0b3$};
\node [anchor=west] at (5,5) {$\labb 0b4$};
\node [anchor=west] at (5,6) {$\labb 0b5$};
\node [anchor=south] at (5,7) {$\vertb$};
\node [anchor=west] at (5,4) {$\labb 0b3$};
\node [anchor=west] at (5,3) {$\labRb 0b2$};

\draw  (7,0) circle (2.0pt);
\draw [fill] (7,1) circle (2.0pt);
\draw [fill] (7,2) circle (2.0pt);
\draw [fill] (8,1) circle (2.0pt);
\draw [fill] (8,2) circle (2.0pt);
\draw [fill] (8,3) circle (2.0pt);
\draw [fill] (8,4) circle (2.0pt);
\draw [fill] (9,3) circle (2.0pt);
\draw [fill] (9,4) circle (2.0pt);
\draw [fill] (9,5) circle (2.0pt);
\draw [fill] (9,6) circle (2.0pt);
\draw  (9,7) circle (2.0pt);
\draw [->,>=stealth,thick,shorten >=2pt,shorten <=2pt] (7,0) -- (7,1);
\draw [->,>=stealth,thick,shorten >=2pt,shorten <=2pt] (7,1) -- (7,2);
\draw [->,>=stealth,thick,shorten >=2pt,shorten <=2pt] (8,1) -- (7,2);
\draw [->,>=stealth,thick,shorten >=2pt,shorten <=2pt] (8,1) -- (8,2);
\draw [->,>=stealth,thick,shorten >=2pt,shorten <=2pt] (8,2) -- (8,3);
\draw [->,>=stealth,thick,shorten >=2pt,shorten <=2pt] (8,3) -- (8,4);
\draw [->,>=stealth,thick,shorten >=2pt,shorten <=2pt] (9,3) -- (8,4);
\draw [->,>=stealth,thick,shorten >=2pt,shorten <=2pt] (9,3) -- (9,4);
\draw [->,>=stealth,thick,shorten >=2pt,shorten <=2pt] (9,4) -- (9,5);
\draw [->,>=stealth,thick,shorten >=2pt,shorten <=2pt] (9,5) -- (9,6);
\draw [->,>=stealth,thick,shorten >=2pt,shorten <=2pt] (9,6) -- (9,7);
\node [anchor=north] at (7,0) {$0$};
\node [anchor=west] at (7,1) {$\labc 0c0$};
\node [anchor=west] at (8,2) {$\labc 0c1$};
\node [anchor=west] at (8,3) {$\labc 0c2$};
\node [anchor=west] at (9,4) {$\labc 0c3$};
\node [anchor=west] at (9,5) {$\labc 0c4$};
\node [anchor=west] at (9,6) {$\labc 0c5$};
\node [anchor=south] at (9,7) {$\vertc$};
\node [anchor=east] at (7,2) {$\labLc 0c1$};
\node [anchor=west] at (8,1) {$\labRc 0c0$};
\node [anchor=west] at (9,3) {$\labRc 0c2$};
\node [anchor=east] at (8,4) {$\labLc 0c3$};

\draw  (11,0) circle (2.0pt);
\draw [fill] (11,1) circle (2.0pt);
\draw [fill] (11,2) circle (2.0pt);
\draw [fill] (12,1) circle (2.0pt);
\draw [fill] (12,2) circle (2.0pt);
\draw [fill] (12,3) circle (2.0pt);
\draw [fill] (13,2) circle (2.0pt);
\draw [fill] (13,3) circle (2.0pt);
\draw [fill] (13,4) circle (2.0pt);
\draw [fill] (14,4) circle (2.0pt);
\draw [fill] (13,5) circle (2.0pt);
\draw [fill] (14,5) circle (2.0pt);
\draw [fill] (14,6) circle (2.0pt);
\draw  (14,7) circle (2.0pt);
\draw [->,>=stealth,thick,shorten >=2pt,shorten <=2pt] (11,0) -- (11,1);
\draw [->,>=stealth,thick,shorten >=2pt,shorten <=2pt] (11,1) -- (11,2);
\draw [->,>=stealth,thick,shorten >=2pt,shorten <=2pt] (12,1) -- (11,2);
\draw [->,>=stealth,thick,shorten >=2pt,shorten <=2pt] (12,1) -- (12,2);
\draw [->,>=stealth,thick,shorten >=2pt,shorten <=2pt] (12,2) -- (12,3);
\draw [->,>=stealth,thick,shorten >=2pt,shorten <=2pt] (13,2) -- (12,3);
\draw [->,>=stealth,thick,shorten >=2pt,shorten <=2pt] (13,2) -- (13,3);
\draw [->,>=stealth,thick,shorten >=2pt,shorten <=2pt] (13,3) -- (13,4);
\draw [->,>=stealth,thick,shorten >=2pt,shorten <=2pt] (13,4) -- (13,5);
\draw [->,>=stealth,thick,shorten >=2pt,shorten <=2pt] (14,4) -- (13,5);
\draw [->,>=stealth,thick,shorten >=2pt,shorten <=2pt] (14,4) -- (14,5);
\draw [->,>=stealth,thick,shorten >=2pt,shorten <=2pt] (14,5) -- (14,6);
\draw [->,>=stealth,thick,shorten >=2pt,shorten <=2pt] (14,6) -- (14,7);
\node [anchor=north] at (11,0) {$0$};
\node [anchor=west] at (11,1) {$\labd 0d0$};
\node [anchor=west] at (12,2) {$\labd 0d1$};
\node [anchor=west] at (13,2) {$\labRd 0d1$};
\node [anchor=east] at (12,3) {$\labLd 0d2$};
\node [anchor=west] at (14,4) {$\labRd 0d3$};
\node [anchor=west] at (14,5) {$\labd 0d4$};
\node [anchor=west] at (14,6) {$\labd 0d5$};
\node [anchor=east] at (13,5) {$\labLd 0d4$};
\node [anchor=south] at (14,7) {$\vertd$};
\node [anchor=east] at (11,2) {$\labLd 0d1$};
\node [anchor=west] at (12,1) {$\labRd 0d0$};
\node [anchor=west] at (13,3) {$\labd 0d2$};
\node [anchor=west] at (13,4) {$\labd 0d3$};

\draw  (16,0) circle (2.0pt);
\draw [fill] (16,1) circle (2.0pt);
\draw [fill] (16,2) circle (2.0pt);
\draw [fill] (17,1) circle (2.0pt);
\draw [fill] (17,2) circle (2.0pt);
\draw [fill] (17,3) circle (2.0pt);
\draw [fill] (18,2) circle (2.0pt);
\draw [fill] (18,3) circle (2.0pt);
\draw [fill] (18,4) circle (2.0pt);
\draw [fill] (19,3) circle (2.0pt);
\draw [fill] (19,4) circle (2.0pt);
\draw [fill] (19,5) circle (2.0pt);
\draw [fill] (20,4) circle (2.0pt);
\draw [fill] (20,5) circle (2.0pt);
\draw [fill] (20,6) circle (2.0pt);
\draw  (20,7) circle (2.0pt);
\draw [->,>=stealth,thick,shorten >=2pt,shorten <=2pt] (16,0) -- (16,1);
\draw [->,>=stealth,thick,shorten >=2pt,shorten <=2pt] (16,1) -- (16,2);
\draw [->,>=stealth,thick,shorten >=2pt,shorten <=2pt] (17,1) -- (16,2);
\draw [->,>=stealth,thick,shorten >=2pt,shorten <=2pt] (17,1) -- (17,2);
\draw [->,>=stealth,thick,shorten >=2pt,shorten <=2pt] (17,2) -- (17,3);
\draw [->,>=stealth,thick,shorten >=2pt,shorten <=2pt] (18,2) -- (17,3);
\draw [->,>=stealth,thick,shorten >=2pt,shorten <=2pt] (18,2) -- (18,3);
\draw [->,>=stealth,thick,shorten >=2pt,shorten <=2pt] (18,3) -- (18,4);
\draw [->,>=stealth,thick,shorten >=2pt,shorten <=2pt] (19,3) -- (18,4);
\draw [->,>=stealth,thick,shorten >=2pt,shorten <=2pt] (19,3) -- (19,4);
\draw [->,>=stealth,thick,shorten >=2pt,shorten <=2pt] (19,4) -- (19,5);
\draw [->,>=stealth,thick,shorten >=2pt,shorten <=2pt] (20,4) -- (19,5);
\draw [->,>=stealth,thick,shorten >=2pt,shorten <=2pt] (20,4) -- (20,5);
\draw [->,>=stealth,thick,shorten >=2pt,shorten <=2pt] (20,5) -- (20,6);
\draw [->,>=stealth,thick,shorten >=2pt,shorten <=2pt] (20,6) -- (20,7);
\node [anchor=north] at (16,0) {$0$};
\node [anchor=west] at (16,1) {$\labt 0t0$};
\node [anchor=west] at (17,2) {$\labt 0t1$};
\node [anchor=west] at (18,2) {$\labRt 0t1$};
\node [anchor=east] at (17,3) {$\labLt 0t2$};
\node [anchor=west] at (20,4) {$\labRt 0t3$};
\node [anchor=west] at (20,5) {$\labt 0t4$};
\node [anchor=west] at (20,6) {$\labt 0t5$};
\node [anchor=east] at (19,5) {$\labLt 0t4$};
\node [anchor=south] at (20,7) {$\vertt$};
\node [anchor=east] at (16,2) {$\labLt 0t1$};
\node [anchor=west] at (17,1) {$\labRt 0t0$};
\node [anchor=west] at (19,4) {$\labt 0t3$};
\node [anchor=east] at (18,4) {$\labLt 0t3$};
\node [anchor=west] at (18,3) {$\labt 0t2$};
\node [anchor=west] at (19,3) {$\labRt 0t2$};
\end{tikzpicture}

\hspace{-.7in}
\begin{tikzpicture}[scale=.8]

\draw  (-2,0) circle (2.0pt);
\draw [fill] (-2,1) circle (2.0pt);
\draw [fill] (-2,2) circle (2.0pt);
\draw [fill] (-1,1) circle (2.0pt);
\draw [fill] (-1,2) circle (2.0pt);
\draw [fill] (-1,3) circle (2.0pt);
\draw [fill] (0,2) circle (2.0pt);
\draw [fill] (0,3) circle (2.0pt);
\draw [fill] (0,4) circle (2.0pt);
\draw [fill] (1,3) circle (2.0pt);
\draw [fill] (1,4) circle (2.0pt);
\draw [fill] (1,5) circle (2.0pt);
\draw [fill] (1,6) circle (2.0pt);
\draw [fill] (2,5) circle (2.0pt);
\draw [fill] (2,6) circle (2.0pt);
\draw  (2,7) circle (2.0pt);
\draw [->,>=stealth,thick,shorten >=2pt,shorten <=2pt] (-2,0) -- (-2,1);
\draw [->,>=stealth,thick,shorten >=2pt,shorten <=2pt] (-2,1) -- (-2,2);
\draw [->,>=stealth,thick,shorten >=2pt,shorten <=2pt] (-1,1) -- (-2,2);
\draw [->,>=stealth,thick,shorten >=2pt,shorten <=2pt] (-1,1) -- (-1,2);
\draw [->,>=stealth,thick,shorten >=2pt,shorten <=2pt] (-1,2) -- (-1,3);
\draw [->,>=stealth,thick,shorten >=2pt,shorten <=2pt] (0,2) -- (-1,3);
\draw [->,>=stealth,thick,shorten >=2pt,shorten <=2pt] (0,2) -- (0,3);
\draw [->,>=stealth,thick,shorten >=2pt,shorten <=2pt] (0,3) -- (0,4);
\draw [->,>=stealth,thick,shorten >=2pt,shorten <=2pt] (1,3) -- (0,4);
\draw [->,>=stealth,thick,shorten >=2pt,shorten <=2pt] (1,3) -- (1,4);
\draw [->,>=stealth,thick,shorten >=2pt,shorten <=2pt] (1,4) -- (1,5);
\draw [->,>=stealth,thick,shorten >=2pt,shorten <=2pt] (1,5) -- (1,6);
\draw [->,>=stealth,thick,shorten >=2pt,shorten <=2pt] (2,5) -- (1,6);
\draw [->,>=stealth,thick,shorten >=2pt,shorten <=2pt] (2,5) -- (2,6);
\draw [->,>=stealth,thick,shorten >=2pt,shorten <=2pt] (2,6) -- (2,7);
\node [anchor=north] at (-2,0) {$1$};
\node [anchor=west] at (-2,1) {$\laba 1a0$};
\node [anchor=east] at (-2,2) {$\labLa 1a1$};
\node [anchor=west] at (-1,1) {$\labRa 1a0$};
\node [anchor=west] at (-1,2) {$\laba 1a1$};
\node [anchor=east] at (-1,3) {$\labLa 1a2$};
\node [anchor=west] at (0,2) {$\labRa 1a1$};
\node [anchor=west] at (0,3) {$\laba 1a2$};
\node [anchor=west] at (1,3) {$\labRa 1a2$};
\node [anchor=west] at (1,4) {$\laba 1a3$};
\node [anchor=east] at (0,4) {$\labLa 1a3$};
\node [anchor=west] at (1,5) {$\laba 1a4$};
\node [anchor=west] at (2,5) {$\labRa 1a4$};
\node [anchor=east] at (1,6) {$\labLa 1a5$};
\node [anchor=west] at (2,6) {$\laba 1a5$};
\node [anchor=south] at (2,7) {$\verta$};

\draw  (3,0) circle (2.0pt);
\draw [fill] (3,1) circle (2.0pt);
\draw [fill] (3,2) circle (2.0pt);
\draw [fill] (4,1) circle (2.0pt);
\draw [fill] (4,2) circle (2.0pt);
\draw [fill] (4,3) circle (2.0pt);
\draw [fill] (4,4) circle (2.0pt);
\draw [fill] (4,5) circle (2.0pt);
\draw [fill] (5,4) circle (2.0pt);
\draw [fill] (5,5) circle (2.0pt);
\draw [fill] (5,6) circle (2.0pt);
\draw [fill] (6,5) circle (2.0pt);
\draw [fill] (6,6) circle (2.0pt);
\draw  (6,7) circle (2.0pt);
\draw [->,>=stealth,thick,shorten >=2pt,shorten <=2pt] (3,0) -- (3,1);
\draw [->,>=stealth,thick,shorten >=2pt,shorten <=2pt] (3,1) -- (3,2);
\draw [->,>=stealth,thick,shorten >=2pt,shorten <=2pt] (4,1) -- (3,2);
\draw [->,>=stealth,thick,shorten >=2pt,shorten <=2pt] (4,1) -- (4,2);
\draw [->,>=stealth,thick,shorten >=2pt,shorten <=2pt] (4,2) -- (4,3);
\draw [->,>=stealth,thick,shorten >=2pt,shorten <=2pt] (4,3) -- (4,4);
\draw [->,>=stealth,thick,shorten >=2pt,shorten <=2pt] (4,4) -- (4,5);
\draw [->,>=stealth,thick,shorten >=2pt,shorten <=2pt] (5,4) -- (4,5);
\draw [->,>=stealth,thick,shorten >=2pt,shorten <=2pt] (5,4) -- (5,5);
\draw [->,>=stealth,thick,shorten >=2pt,shorten <=2pt] (5,5) -- (5,6);
\draw [->,>=stealth,thick,shorten >=2pt,shorten <=2pt] (6,5) -- (5,6);
\draw [->,>=stealth,thick,shorten >=2pt,shorten <=2pt] (6,5) -- (6,6);
\draw [->,>=stealth,thick,shorten >=2pt,shorten <=2pt] (6,6) -- (6,7);
\node [anchor=north] at (3,0) {$1$};
\node [anchor=west] at (3,1) {$\labb 1b0$};
\node [anchor=east] at (3,2) {$\labLb 1b1$};
\node [anchor=west] at (4,1) {$\labRb 1b0$};
\node [anchor=west] at (4,2) {$\labb 1b1$};
\node [anchor=west] at (4,3) {$\labb 1b2$};
\node [anchor=west] at (4,4) {$\labb 1b3$};
\node [anchor=west] at (5,5) {$\labb 1b4$};
\node [anchor=west] at (6,5) {$\labRb 1b4$};
\node [anchor=east] at (4,5) {$\labLb 1b4$};
\node [anchor=south] at (6,7) {$\vertb$};
\node [anchor=west] at (5,4) {$\labRb 1b3$};
\node [anchor=west] at (6,6) {$\labb 1b5$};
\node [anchor=east] at (5,6) {$\labLb 1b5$};

\draw  (7,0) circle (2.0pt);
\draw [fill] (7,1) circle (2.0pt);
\draw [fill] (7,2) circle (2.0pt);
\draw [fill] (7,3) circle (2.0pt);
\draw [fill] (8,2) circle (2.0pt);
\draw [fill] (8,3) circle (2.0pt);
\draw [fill] (8,4) circle (2.0pt);
\draw [fill] (8,5) circle (2.0pt);
\draw [fill] (9,4) circle (2.0pt);
\draw [fill] (9,5) circle (2.0pt);
\draw [fill] (9,6) circle (2.0pt);
\draw [fill] (10,5) circle (2.0pt);
\draw [fill] (10,6) circle (2.0pt);
\draw  (10,7) circle (2.0pt);
\draw [->,>=stealth,thick,shorten >=2pt,shorten <=2pt] (7,0) -- (7,1);
\draw [->,>=stealth,thick,shorten >=2pt,shorten <=2pt] (7,1) -- (7,2);
\draw [->,>=stealth,thick,shorten >=2pt,shorten <=2pt] (7,2) -- (7,3);
\draw [->,>=stealth,thick,shorten >=2pt,shorten <=2pt] (8,2) -- (7,3);
\draw [->,>=stealth,thick,shorten >=2pt,shorten <=2pt] (8,2) -- (8,3);
\draw [->,>=stealth,thick,shorten >=2pt,shorten <=2pt] (8,3) -- (8,4);
\draw [->,>=stealth,thick,shorten >=2pt,shorten <=2pt] (8,4) -- (8,5);
\draw [->,>=stealth,thick,shorten >=2pt,shorten <=2pt] (9,4) -- (8,5);
\draw [->,>=stealth,thick,shorten >=2pt,shorten <=2pt] (9,4) -- (9,5);
\draw [->,>=stealth,thick,shorten >=2pt,shorten <=2pt] (9,5) -- (9,6);
\draw [->,>=stealth,thick,shorten >=2pt,shorten <=2pt] (10,5) -- (9,6);
\draw [->,>=stealth,thick,shorten >=2pt,shorten <=2pt] (10,5) -- (10,6);
\draw [->,>=stealth,thick,shorten >=2pt,shorten <=2pt] (10,6) -- (10,7);
\node [anchor=north] at (7,0) {$1$};
\node [anchor=west] at (7,1) {$\labc 1c0$};
\node [anchor=west] at (8,2) {$\labRc 1c1$};
\node [anchor=west] at (8,3) {$\labc 1c2$};
\node [anchor=east] at (7,3) {$\labLc 1c2$};
\node [anchor=west] at (9,4) {$\labRc 1c3$};
\node [anchor=west] at (9,5) {$\labc 1c4$};
\node [anchor=west] at (10,5) {$\labRc 1c4$};
\node [anchor=east] at (8,5) {$\labLc 1c4$};
\node [anchor=east] at (9,6) {$\labLc 1c5$};
\node [anchor=west] at (10,6) {$\labc 1c5$};
\node [anchor=south] at (10,7) {$\vertc$};
\node [anchor=west] at (7,2) {$\labc 1c1$};
\node [anchor=west] at (8,4) {$\labc 1c3$};

\draw  (11,0) circle (2.0pt);
\draw [fill] (11,1) circle (2.0pt);
\draw [fill] (11,2) circle (2.0pt);
\draw [fill] (11,3) circle (2.0pt);
\draw [fill] (11,4) circle (2.0pt);
\draw [fill] (12,3) circle (2.0pt);
\draw [fill] (12,4) circle (2.0pt);
\draw [fill] (12,5) circle (2.0pt);
\draw [fill] (12,6) circle (2.0pt);
\draw [fill] (13,5) circle (2.0pt);
\draw [fill] (13,6) circle (2.0pt);
\draw  (13,7) circle (2.0pt);
\draw [->,>=stealth,thick,shorten >=2pt,shorten <=2pt] (11,0) -- (11,1);
\draw [->,>=stealth,thick,shorten >=2pt,shorten <=2pt] (11,1) -- (11,2);
\draw [->,>=stealth,thick,shorten >=2pt,shorten <=2pt] (11,2) -- (11,3);
\draw [->,>=stealth,thick,shorten >=2pt,shorten <=2pt] (11,3) -- (11,4);
\draw [->,>=stealth,thick,shorten >=2pt,shorten <=2pt] (12,3) -- (11,4);
\draw [->,>=stealth,thick,shorten >=2pt,shorten <=2pt] (12,3) -- (12,4);
\draw [->,>=stealth,thick,shorten >=2pt,shorten <=2pt] (12,4) -- (12,5);
\draw [->,>=stealth,thick,shorten >=2pt,shorten <=2pt] (12,5) -- (12,6);
\draw [->,>=stealth,thick,shorten >=2pt,shorten <=2pt] (13,5) -- (12,6);
\draw [->,>=stealth,thick,shorten >=2pt,shorten <=2pt] (13,5) -- (13,6);
\draw [->,>=stealth,thick,shorten >=2pt,shorten <=2pt] (13,6) -- (13,7);
\node [anchor=north] at (11,0) {$1$};
\node [anchor=west] at (11,1) {$\labd 1d0$};
\node [anchor=west] at (11,2) {$\labd 1d1$};
\node [anchor=west] at (11,3) {$\labd 1d2$};
\node [anchor=west] at (12,3) {$\labRd 1d2$};
\node [anchor=west] at (12,4) {$\labd 1d3$};
\node [anchor=east] at (11,4) {$\labLd 1d3$};
\node [anchor=west] at (12,5) {$\labd 1d4$};
\node [anchor=west] at (13,5) {$\labRd 1d4$};
\node [anchor=west] at (13,6) {$\labd 1d5$};
\node [anchor=east] at (12,6) {$\labLd 1d5$};
\node [anchor=south] at (13,7) {$\vertd$};

\draw  (14,0) circle (2.0pt);
\draw [fill] (14,1) circle (2.0pt);
\draw [fill] (14,2) circle (2.0pt);
\draw [fill] (15,1) circle (2.0pt);
\draw [fill] (15,2) circle (2.0pt);
\draw [fill] (15,3) circle (2.0pt);
\draw [fill] (16,2) circle (2.0pt);
\draw [fill] (16,3) circle (2.0pt);
\draw [fill] (16,4) circle (2.0pt);
\draw [fill] (17,3) circle (2.0pt);
\draw [fill] (17,4) circle (2.0pt);
\draw [fill] (17,5) circle (2.0pt);
\draw [fill] (18,4) circle (2.0pt);
\draw [fill] (18,5) circle (2.0pt);
\draw [fill] (18,6) circle (2.0pt);
\draw [fill] (19,5) circle (2.0pt);
\draw [fill] (19,6) circle (2.0pt);
\draw  (19,7) circle (2.0pt);
\draw [->,>=stealth,thick,shorten >=2pt,shorten <=2pt] (14,0) -- (14,1);
\draw [->,>=stealth,thick,shorten >=2pt,shorten <=2pt] (14,1) -- (14,2);
\draw [->,>=stealth,thick,shorten >=2pt,shorten <=2pt] (15,1) -- (14,2);
\draw [->,>=stealth,thick,shorten >=2pt,shorten <=2pt] (15,1) -- (15,2);
\draw [->,>=stealth,thick,shorten >=2pt,shorten <=2pt] (15,2) -- (15,3);
\draw [->,>=stealth,thick,shorten >=2pt,shorten <=2pt] (16,2) -- (15,3);
\draw [->,>=stealth,thick,shorten >=2pt,shorten <=2pt] (16,2) -- (16,3);
\draw [->,>=stealth,thick,shorten >=2pt,shorten <=2pt] (16,3) -- (16,4);
\draw [->,>=stealth,thick,shorten >=2pt,shorten <=2pt] (17,3) -- (16,4);
\draw [->,>=stealth,thick,shorten >=2pt,shorten <=2pt] (17,3) -- (17,4);
\draw [->,>=stealth,thick,shorten >=2pt,shorten <=2pt] (17,4) -- (17,5);
\draw [->,>=stealth,thick,shorten >=2pt,shorten <=2pt] (18,4) -- (17,5);
\draw [->,>=stealth,thick,shorten >=2pt,shorten <=2pt] (18,4) -- (18,5);
\draw [->,>=stealth,thick,shorten >=2pt,shorten <=2pt] (18,5) -- (18,6);
\draw [->,>=stealth,thick,shorten >=2pt,shorten <=2pt] (19,5) -- (18,6);
\draw [->,>=stealth,thick,shorten >=2pt,shorten <=2pt] (19,5) -- (19,6);
\draw [->,>=stealth,thick,shorten >=2pt,shorten <=2pt] (19,6) -- (19,7);
\node [anchor=north] at (14,0) {$1$};
\node [anchor=west] at (14,1) {$\labt 1t0$};
\node [anchor=west] at (15,2) {$\labt 1t1$};
\node [anchor=west] at (16,2) {$\labRt 1t1$};
\node [anchor=east] at (15,3) {$\labLt 1t2$};
\node [anchor=west] at (18,4) {$\labRt 1t3$};
\node [anchor=west] at (18,5) {$\labt 1t4$};
\node [anchor=east] at (18,6) {$\labLt 1t5$};
\node [anchor=west] at (19,5) {$\labRt 1t4$};
\node [anchor=west] at (19,6) {$\labt 1t5$};
\node [anchor=east] at (17,5) {$\labLt 1t4$};
\node [anchor=south] at (19,7) {$\vertt$};
\node [anchor=east] at (14,2) {$\labLt 1t1$};
\node [anchor=west] at (15,1) {$\labRt 1t0$};
\node [anchor=west] at (17,4) {$\labt 1t3$};
\node [anchor=east] at (16,4) {$\labLt 1t3$};
\node [anchor=west] at (16,3) {$\labt 1t2$};
\node [anchor=west] at (17,3) {$\labRt 1t2$};

\end{tikzpicture}

\hspace{-.8in}
\begin{tikzpicture}[scale=.8]

\def\x{-14}
\draw  (14+\x,0) circle (2.0pt);
\draw [fill] (14+\x,1) circle (2.0pt);
\draw [fill] (14+\x,2) circle (2.0pt);
\draw [fill] (15+\x,1) circle (2.0pt);
\draw [fill] (15+\x,2) circle (2.0pt);
\draw [fill] (15+\x,3) circle (2.0pt);
\draw [fill] (16+\x,2) circle (2.0pt);
\draw [fill] (16+\x,3) circle (2.0pt);
\draw [fill] (16+\x,4) circle (2.0pt);
\draw [fill] (17+\x,3) circle (2.0pt);
\draw [fill] (17+\x,4) circle (2.0pt);
\draw [fill] (17+\x,5) circle (2.0pt);
\draw [fill] (18+\x,4) circle (2.0pt);
\draw [fill] (18+\x,5) circle (2.0pt);
\draw [fill] (18+\x,6) circle (2.0pt);
\draw [fill] (19+\x,5) circle (2.0pt);
\draw [fill] (19+\x,6) circle (2.0pt);
\draw  (19+\x,7) circle (2.0pt);
\draw [->,>=stealth,thick,shorten >=2pt,shorten <=2pt] (14+\x,0) -- (14+\x,1);
\draw [->,>=stealth,thick,shorten >=2pt,shorten <=2pt] (14+\x,1) -- (14+\x,2);
\draw [->,>=stealth,thick,shorten >=2pt,shorten <=2pt] (15+\x,1) -- (14+\x,2);
\draw [->,>=stealth,thick,shorten >=2pt,shorten <=2pt] (15+\x,1) -- (15+\x,2);
\draw [->,>=stealth,thick,shorten >=2pt,shorten <=2pt] (15+\x,2) -- (15+\x,3);
\draw [->,>=stealth,thick,shorten >=2pt,shorten <=2pt] (16+\x,2) -- (15+\x,3);
\draw [->,>=stealth,thick,shorten >=2pt,shorten <=2pt] (16+\x,2) -- (16+\x,3);
\draw [->,>=stealth,thick,shorten >=2pt,shorten <=2pt] (16+\x,3) -- (16+\x,4);
\draw [->,>=stealth,thick,shorten >=2pt,shorten <=2pt] (17+\x,3) -- (16+\x,4);
\draw [->,>=stealth,thick,shorten >=2pt,shorten <=2pt] (17+\x,3) -- (17+\x,4);
\draw [->,>=stealth,thick,shorten >=2pt,shorten <=2pt] (17+\x,4) -- (17+\x,5);
\draw [->,>=stealth,thick,shorten >=2pt,shorten <=2pt] (18+\x,4) -- (17+\x,5);
\draw [->,>=stealth,thick,shorten >=2pt,shorten <=2pt] (18+\x,4) -- (18+\x,5);
\draw [->,>=stealth,thick,shorten >=2pt,shorten <=2pt] (18+\x,5) -- (18+\x,6);
\draw [->,>=stealth,thick,shorten >=2pt,shorten <=2pt] (19+\x,5) -- (18+\x,6);
\draw [->,>=stealth,thick,shorten >=2pt,shorten <=2pt] (19+\x,5) -- (19+\x,6);
\draw [->,>=stealth,thick,shorten >=2pt,shorten <=2pt] (19+\x,6) -- (19+\x,7);
\node [anchor=north] at (14+\x,0) {$2$};
\node [anchor=west] at (14+\x,1) {$\laba 2a0$};
\node [anchor=west] at (15+\x,2) {$\laba 2a1$};
\node [anchor=west] at (16+\x,2) {$\labRa 2a1$};
\node [anchor=east] at (15+\x,3) {$\labLa 2a2$};
\node [anchor=west] at (18+\x,4) {$\labRa 2a3$};
\node [anchor=west] at (18+\x,5) {$\laba 2a4$};
\node [anchor=east] at (18+\x,6) {$\labLa 2a5$};
\node [anchor=east] at (17+\x,5) {$\labLa 2a4$};
\node [anchor=west] at (19+\x,5) {$\labRa 2a4$};
\node [anchor=west] at (19+\x,6) {$\laba 2a5$};
\node [anchor=south] at (19+\x,7) {$\verta$};
\node [anchor=east] at (14+\x,2) {$\labLa 2a1$};
\node [anchor=west] at (15+\x,1) {$\labRa 2a0$};
\node [anchor=west] at (17+\x,4) {$\laba 2a3$};
\node [anchor=east] at (16+\x,4) {$\labLa 2a3$};
\node [anchor=west] at (16+\x,3) {$\laba 2a2$};
\node [anchor=west] at (17+\x,3) {$\labRa 2a2$};

\def\x{-9.5}
\draw  (14+\x,0) circle (2.0pt);
\draw [fill] (14+\x,1) circle (2.0pt);
\draw [fill] (14+\x,2) circle (2.0pt);
\draw [fill] (15+\x,1) circle (2.0pt);
\draw [fill] (15+\x,2) circle (2.0pt);
\draw [fill] (15+\x,3) circle (2.0pt);
\draw [fill] (16+\x,2) circle (2.0pt);
\draw [fill] (16+\x,3) circle (2.0pt);
\draw [fill] (16+\x,4) circle (2.0pt);
\draw [fill] (17+\x,3) circle (2.0pt);
\draw [fill] (17+\x,4) circle (2.0pt);
\draw [fill] (17+\x,5) circle (2.0pt);
\draw [fill] (18+\x,4) circle (2.0pt);
\draw [fill] (18+\x,5) circle (2.0pt);
\draw [fill] (18+\x,6) circle (2.0pt);
\draw [fill] (19+\x,5) circle (2.0pt);
\draw [fill] (19+\x,6) circle (2.0pt);
\draw  (19+\x,7) circle (2.0pt);
\draw [->,>=stealth,thick,shorten >=2pt,shorten <=2pt] (14+\x,0) -- (14+\x,1);
\draw [->,>=stealth,thick,shorten >=2pt,shorten <=2pt] (14+\x,1) -- (14+\x,2);
\draw [->,>=stealth,thick,shorten >=2pt,shorten <=2pt] (15+\x,1) -- (14+\x,2);
\draw [->,>=stealth,thick,shorten >=2pt,shorten <=2pt] (15+\x,1) -- (15+\x,2);
\draw [->,>=stealth,thick,shorten >=2pt,shorten <=2pt] (15+\x,2) -- (15+\x,3);
\draw [->,>=stealth,thick,shorten >=2pt,shorten <=2pt] (16+\x,2) -- (15+\x,3);
\draw [->,>=stealth,thick,shorten >=2pt,shorten <=2pt] (16+\x,2) -- (16+\x,3);
\draw [->,>=stealth,thick,shorten >=2pt,shorten <=2pt] (16+\x,3) -- (16+\x,4);
\draw [->,>=stealth,thick,shorten >=2pt,shorten <=2pt] (17+\x,3) -- (16+\x,4);
\draw [->,>=stealth,thick,shorten >=2pt,shorten <=2pt] (17+\x,3) -- (17+\x,4);
\draw [->,>=stealth,thick,shorten >=2pt,shorten <=2pt] (17+\x,4) -- (17+\x,5);
\draw [->,>=stealth,thick,shorten >=2pt,shorten <=2pt] (18+\x,4) -- (17+\x,5);
\draw [->,>=stealth,thick,shorten >=2pt,shorten <=2pt] (18+\x,4) -- (18+\x,5);
\draw [->,>=stealth,thick,shorten >=2pt,shorten <=2pt] (18+\x,5) -- (18+\x,6);
\draw [->,>=stealth,thick,shorten >=2pt,shorten <=2pt] (19+\x,5) -- (18+\x,6);
\draw [->,>=stealth,thick,shorten >=2pt,shorten <=2pt] (19+\x,5) -- (19+\x,6);
\draw [->,>=stealth,thick,shorten >=2pt,shorten <=2pt] (19+\x,6) -- (19+\x,7);
\node [anchor=north] at (14+\x,0) {$2$};
\node [anchor=west] at (14+\x,1) {$\labb 2b0$};
\node [anchor=west] at (15+\x,2) {$\labb 2b1$};
\node [anchor=west] at (16+\x,2) {$\labRb 2b1$};
\node [anchor=east] at (15+\x,3) {$\labLb 2b2$};
\node [anchor=west] at (18+\x,4) {$\labRb 2b3$};
\node [anchor=west] at (19+\x,5) {$\labRb 2b4$};
\node [anchor=west] at (18+\x,5) {$\labb 2b4$};
\node [anchor=west] at (19+\x,6) {$\labb 2b5$};
\node [anchor=east] at (17+\x,5) {$\labLb 2b4$};
\node [anchor=east] at (18+\x,6) {$\labLb 2b5$};
\node [anchor=south] at (19+\x,7) {$\vertb$};
\node [anchor=east] at (14+\x,2) {$\labLb 2b1$};
\node [anchor=west] at (15+\x,1) {$\labRb 2b0$};
\node [anchor=west] at (17+\x,4) {$\labb 2b3$};
\node [anchor=east] at (16+\x,4) {$\labLb 2b3$};
\node [anchor=west] at (16+\x,3) {$\labb 2b2$};
\node [anchor=west] at (17+\x,3) {$\labRb 2b2$};

\def\x{-5}
\draw  (14+\x,0) circle (2.0pt);
\draw [fill] (14+\x,1) circle (2.0pt);
\draw [fill] (14+\x,2) circle (2.0pt);
\draw [fill] (15+\x,1) circle (2.0pt);
\draw [fill] (15+\x,2) circle (2.0pt);
\draw [fill] (15+\x,3) circle (2.0pt);
\draw [fill] (16+\x,2) circle (2.0pt);
\draw [fill] (16+\x,3) circle (2.0pt);
\draw [fill] (16+\x,4) circle (2.0pt);
\draw [fill] (17+\x,3) circle (2.0pt);
\draw [fill] (17+\x,4) circle (2.0pt);
\draw [fill] (17+\x,5) circle (2.0pt);
\draw [fill] (18+\x,4) circle (2.0pt);
\draw [fill] (18+\x,5) circle (2.0pt);
\draw [fill] (18+\x,6) circle (2.0pt);
\draw [fill] (19+\x,5) circle (2.0pt);
\draw [fill] (19+\x,6) circle (2.0pt);
\draw  (19+\x,7) circle (2.0pt);
\draw [->,>=stealth,thick,shorten >=2pt,shorten <=2pt] (14+\x,0) -- (14+\x,1);
\draw [->,>=stealth,thick,shorten >=2pt,shorten <=2pt] (14+\x,1) -- (14+\x,2);
\draw [->,>=stealth,thick,shorten >=2pt,shorten <=2pt] (15+\x,1) -- (14+\x,2);
\draw [->,>=stealth,thick,shorten >=2pt,shorten <=2pt] (15+\x,1) -- (15+\x,2);
\draw [->,>=stealth,thick,shorten >=2pt,shorten <=2pt] (15+\x,2) -- (15+\x,3);
\draw [->,>=stealth,thick,shorten >=2pt,shorten <=2pt] (16+\x,2) -- (15+\x,3);
\draw [->,>=stealth,thick,shorten >=2pt,shorten <=2pt] (16+\x,2) -- (16+\x,3);
\draw [->,>=stealth,thick,shorten >=2pt,shorten <=2pt] (16+\x,3) -- (16+\x,4);
\draw [->,>=stealth,thick,shorten >=2pt,shorten <=2pt] (17+\x,3) -- (16+\x,4);
\draw [->,>=stealth,thick,shorten >=2pt,shorten <=2pt] (17+\x,3) -- (17+\x,4);
\draw [->,>=stealth,thick,shorten >=2pt,shorten <=2pt] (17+\x,4) -- (17+\x,5);
\draw [->,>=stealth,thick,shorten >=2pt,shorten <=2pt] (18+\x,4) -- (17+\x,5);
\draw [->,>=stealth,thick,shorten >=2pt,shorten <=2pt] (18+\x,4) -- (18+\x,5);
\draw [->,>=stealth,thick,shorten >=2pt,shorten <=2pt] (18+\x,5) -- (18+\x,6);
\draw [->,>=stealth,thick,shorten >=2pt,shorten <=2pt] (19+\x,5) -- (18+\x,6);
\draw [->,>=stealth,thick,shorten >=2pt,shorten <=2pt] (19+\x,5) -- (19+\x,6);
\draw [->,>=stealth,thick,shorten >=2pt,shorten <=2pt] (19+\x,6) -- (19+\x,7);
\node [anchor=north] at (14+\x,0) {$2$};
\node [anchor=west] at (14+\x,1) {$\labc 2c0$};
\node [anchor=west] at (15+\x,2) {$\labc 2c1$};
\node [anchor=west] at (16+\x,2) {$\labRc 2c1$};
\node [anchor=east] at (15+\x,3) {$\labLc 2c2$};
\node [anchor=west] at (18+\x,4) {$\labRc 2c3$};
\node [anchor=west] at (19+\x,5) {$\labRc 2c4$};
\node [anchor=west] at (18+\x,5) {$\labc 2c4$};
\node [anchor=west] at (19+\x,6) {$\labc 2c5$};
\node [anchor=east] at (17+\x,5) {$\labLc 2c4$};
\node [anchor=east] at (18+\x,6) {$\labLc 2c5$};
\node [anchor=south] at (19+\x,7) {$\vertc$};
\node [anchor=east] at (14+\x,2) {$\labLc 2c1$};
\node [anchor=west] at (15+\x,1) {$\labRc 2c0$};
\node [anchor=west] at (17+\x,4) {$\labc 2c3$};
\node [anchor=east] at (16+\x,4) {$\labLc 2c3$};
\node [anchor=west] at (16+\x,3) {$\labc 2c2$};
\node [anchor=west] at (17+\x,3) {$\labRc 2c2$};

\def\x{-0.5}
\draw  (14+\x,0) circle (2.0pt);
\draw [fill] (14+\x,1) circle (2.0pt);
\draw [fill] (14+\x,2) circle (2.0pt);
\draw [fill] (15+\x,1) circle (2.0pt);
\draw [fill] (15+\x,2) circle (2.0pt);
\draw [fill] (15+\x,3) circle (2.0pt);
\draw [fill] (16+\x,2) circle (2.0pt);
\draw [fill] (16+\x,3) circle (2.0pt);
\draw [fill] (16+\x,4) circle (2.0pt);
\draw [fill] (17+\x,3) circle (2.0pt);
\draw [fill] (17+\x,4) circle (2.0pt);
\draw [fill] (17+\x,5) circle (2.0pt);
\draw [fill] (18+\x,4) circle (2.0pt);
\draw [fill] (18+\x,5) circle (2.0pt);
\draw [fill] (18+\x,6) circle (2.0pt);
\draw [fill] (19+\x,5) circle (2.0pt);
\draw [fill] (19+\x,6) circle (2.0pt);
\draw  (19+\x,7) circle (2.0pt);
\draw [->,>=stealth,thick,shorten >=2pt,shorten <=2pt] (14+\x,0) -- (14+\x,1);
\draw [->,>=stealth,thick,shorten >=2pt,shorten <=2pt] (14+\x,1) -- (14+\x,2);
\draw [->,>=stealth,thick,shorten >=2pt,shorten <=2pt] (15+\x,1) -- (14+\x,2);
\draw [->,>=stealth,thick,shorten >=2pt,shorten <=2pt] (15+\x,1) -- (15+\x,2);
\draw [->,>=stealth,thick,shorten >=2pt,shorten <=2pt] (15+\x,2) -- (15+\x,3);
\draw [->,>=stealth,thick,shorten >=2pt,shorten <=2pt] (16+\x,2) -- (15+\x,3);
\draw [->,>=stealth,thick,shorten >=2pt,shorten <=2pt] (16+\x,2) -- (16+\x,3);
\draw [->,>=stealth,thick,shorten >=2pt,shorten <=2pt] (16+\x,3) -- (16+\x,4);
\draw [->,>=stealth,thick,shorten >=2pt,shorten <=2pt] (17+\x,3) -- (16+\x,4);
\draw [->,>=stealth,thick,shorten >=2pt,shorten <=2pt] (17+\x,3) -- (17+\x,4);
\draw [->,>=stealth,thick,shorten >=2pt,shorten <=2pt] (17+\x,4) -- (17+\x,5);
\draw [->,>=stealth,thick,shorten >=2pt,shorten <=2pt] (18+\x,4) -- (17+\x,5);
\draw [->,>=stealth,thick,shorten >=2pt,shorten <=2pt] (18+\x,4) -- (18+\x,5);
\draw [->,>=stealth,thick,shorten >=2pt,shorten <=2pt] (18+\x,5) -- (18+\x,6);
\draw [->,>=stealth,thick,shorten >=2pt,shorten <=2pt] (19+\x,5) -- (18+\x,6);
\draw [->,>=stealth,thick,shorten >=2pt,shorten <=2pt] (19+\x,5) -- (19+\x,6);
\draw [->,>=stealth,thick,shorten >=2pt,shorten <=2pt] (19+\x,6) -- (19+\x,7);
\node [anchor=north] at (14+\x,0) {$2$};
\node [anchor=west] at (14+\x,1) {$\labd 2d0$};
\node [anchor=west] at (15+\x,2) {$\labd 2d1$};
\node [anchor=west] at (16+\x,2) {$\labRd 2d1$};
\node [anchor=east] at (15+\x,3) {$\labLd 2d2$};
\node [anchor=west] at (18+\x,4) {$\labRd 2d3$};
\node [anchor=west] at (19+\x,5) {$\labRd 2d4$};
\node [anchor=west] at (18+\x,5) {$\labd 2d4$};
\node [anchor=west] at (19+\x,6) {$\labd 2d5$};
\node [anchor=east] at (17+\x,5) {$\labLd 2d4$};
\node [anchor=east] at (18+\x,6) {$\labLd 2d5$};
\node [anchor=south] at (19+\x,7) {$\vertd$};
\node [anchor=east] at (14+\x,2) {$\labLd 2d1$};
\node [anchor=west] at (15+\x,1) {$\labRd 2d0$};
\node [anchor=west] at (17+\x,4) {$\labd 2d3$};
\node [anchor=east] at (16+\x,4) {$\labLd 2d3$};
\node [anchor=west] at (16+\x,3) {$\labd 2d2$};
\node [anchor=west] at (17+\x,3) {$\labRd 2d2$};

\draw  (20.5,0) circle (2.0pt);
\draw [fill] (20.5,1) circle (2.0pt);
\draw [fill] (20.5,2) circle (2.0pt);
\draw [fill] (20.5,3) circle (2.0pt);
\draw [fill] (20.5,4) circle (2.0pt);
\draw [fill] (20.5,5) circle (2.0pt);
\draw [fill] (20.5,6) circle (2.0pt);
\draw [fill] (21.5,5) circle (2.0pt);
\draw [fill] (21.5,6) circle (2.0pt);
\draw  (21.5,7) circle (2.0pt);
\draw [->,>=stealth,thick,shorten >=2pt,shorten <=2pt] (20.5,0) -- (20.5,1);
\draw [->,>=stealth,thick,shorten >=2pt,shorten <=2pt] (20.5,1) -- (20.5,2);
\draw [->,>=stealth,thick,shorten >=2pt,shorten <=2pt] (20.5,2) -- (20.5,3);
\draw [->,>=stealth,thick,shorten >=2pt,shorten <=2pt] (20.5,3) -- (20.5,4);
\draw [->,>=stealth,thick,shorten >=2pt,shorten <=2pt] (20.5,4) -- (20.5,5);
\draw [->,>=stealth,thick,shorten >=2pt,shorten <=2pt] (20.5,5) -- (20.5,6);
\draw [->,>=stealth,thick,shorten >=2pt,shorten <=2pt] (21.5,5) -- (20.5,6);
\draw [->,>=stealth,thick,shorten >=2pt,shorten <=2pt] (21.5,5) -- (21.5,6);
\draw [->,>=stealth,thick,shorten >=2pt,shorten <=2pt] (21.5,6) -- (21.5,7);
\node [anchor=north] at (20.5,0) {$2$};
\node [anchor=west] at (20.5,1) {$\labt 2t0$};
\node [anchor=west] at (20.5,2) {$\labt 2t1$};
\node [anchor=west] at (20.5,3) {$\labt 2t2$};
\node [anchor=west] at (20.5,4) {$\labt 2t3$};
\node [anchor=west] at (20.5,5) {$\labt 2t4$};
\node [anchor=west] at (21.5,6) {$\labt 2t5$};
\node [anchor=east] at (20.6,6) {$\labLt 2t5$};
\node [anchor=west] at (21.5,5) {$\labRt 2t4$};
\node [anchor=south] at (21.5,7) {$\vertt$};
\end{tikzpicture}

For $\varA\in\{0,1,2\}$ and $\varR \in \{\verta,\vertb,\vertc,\vertd,\vertt\}$,
 let $\bbP_{\varA,\varR}$
denote the oriented path from $\varA$ to $\varR$ 
(pictured above and on the previous page).  
Thus $\bbH$ can be roughly pictured as

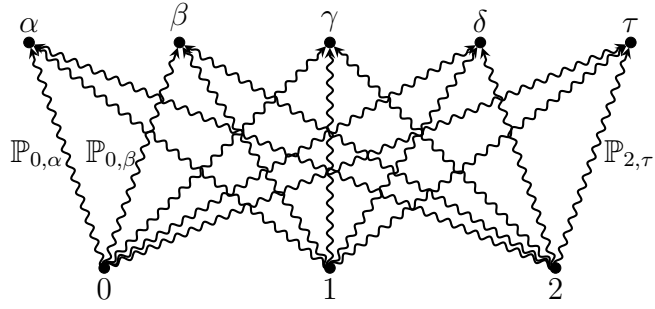
\begin{center}
\begin{tikzpicture}

\draw [fill] (1,0) circle (2.0pt);
\draw [fill] (4,0) circle (2.0pt);
\draw [fill] (7,0) circle (2.0pt);

\draw [fill] (0,3) circle (2.0pt);
\draw [fill] (2,3) circle (2.0pt);
\draw [fill] (4,3) circle (2.0pt);
\draw [fill] (6,3) circle (2.0pt);
\draw [fill] (8,3) circle (2.0pt);

\node [anchor=north] at (1,0) {$0$};
\node [anchor=north] at (4,0) {$1$};
\node [anchor=north] at (7,0) {$2$};
\node [anchor=south] at (0,3) {$\verta$};
\node [anchor=south] at (2,3) {$\vertb$};
\node [anchor=south] at (4,3) {$\vertc$};
\node [anchor=south] at (6,3) {$\vertd$};
\node [anchor=south] at (8,3) {$\vertt$};

\draw [->,snake=snake,segment amplitude=.4mm,segment length=2mm,line after snake=4pt,>=stealth,thick,shorten >=2pt] 
(1,0) -- (0,3);
\draw [->,snake=snake,segment amplitude=.4mm,segment length=2mm,line after snake=4pt,>=stealth,thick,shorten >=2pt] 
(4,0) -- (0,3);
\draw [->,snake=snake,segment amplitude=.4mm,segment length=2mm,line after snake=4pt,>=stealth,thick,shorten >=2pt] 
(7,0) -- (0,3);
\draw [->,snake=snake,segment amplitude=.4mm,segment length=2mm,line after snake=4pt,>=stealth,thick,shorten >=2pt] 
(1,0) -- (2,3);
\draw [->,snake=snake,segment amplitude=.4mm,segment length=2mm,line after snake=4pt,>=stealth,thick,shorten >=2pt] 
(4,0) -- (2,3);
\draw [->,snake=snake,segment amplitude=.4mm,segment length=2mm,line after snake=4pt,>=stealth,thick,shorten >=2pt] 
(7,0) -- (2,3);
\draw [->,snake=snake,segment amplitude=.4mm,segment length=2mm,line after snake=4pt,>=stealth,thick,shorten >=2pt] 
(1,0) -- (4,3);
\draw [->,snake=snake,segment amplitude=.4mm,segment length=2mm,line after snake=4pt,>=stealth,thick,shorten >=2pt] 
(4,0) -- (4,3);
\draw [->,snake=snake,segment amplitude=.4mm,segment length=2mm,line after snake=4pt,>=stealth,thick,shorten >=2pt] 
(7,0) -- (4,3);
\draw [->,snake=snake,segment amplitude=.4mm,segment length=2mm,line after snake=4pt,>=stealth,thick,shorten >=2pt] 
(1,0) -- (6,3);
\draw [->,snake=snake,segment amplitude=.4mm,segment length=2mm,line after snake=4pt,>=stealth,thick,shorten >=2pt] 
(4,0) -- (6,3);
\draw [->,snake=snake,segment amplitude=.4mm,segment length=2mm,line after snake=4pt,>=stealth,thick,shorten >=2pt] 
(7,0) -- (6,3);
\draw [->,snake=snake,segment amplitude=.4mm,segment length=2mm,line after snake=4pt,>=stealth,thick,shorten >=2pt] 
(1,0) -- (8,3);
\draw [->,snake=snake,segment amplitude=.4mm,segment length=2mm,line after snake=4pt,>=stealth,thick,shorten >=2pt] 
(4,0) -- (8,3);
\draw [->,snake=snake,segment amplitude=.4mm,segment length=2mm,line after snake=4pt,>=stealth,thick,shorten >=2pt] 
(7,0) -- (8,3);

\node [anchor=east] at (0.6,1.5) {$\bbP_{0,\verta}$};
\node  at (1.1,1.5) {$\bbP_{0,\vertb}$};
\node [anchor=west] at (7.5,1.5) {$\bbP_{2,\vertt}$};

\end{tikzpicture}
\end{center}

$\bbH$ is core, because the template $\bbA$ on which it is based is core
\cite[Corollary~4.2]{bdjn}, and has a 3-ary idempotent cyclic polymorphism 
$\phi'$ extending $\phi$, because both $\bbA$ and the digraph
$\bullet{\ra}\bullet{\leftarrow}\bullet{\ra}\bullet$ have one
 \cite[Theorem~5.1]{bdjn}.
I will partially describe $\phi'$ in section~\ref{list-sec}.

\subsection{The instance $\bbG$} \label{sec-G}

Let $\bbQ_1,\ldots,\bbQ_4$ denote the following oriented paths:

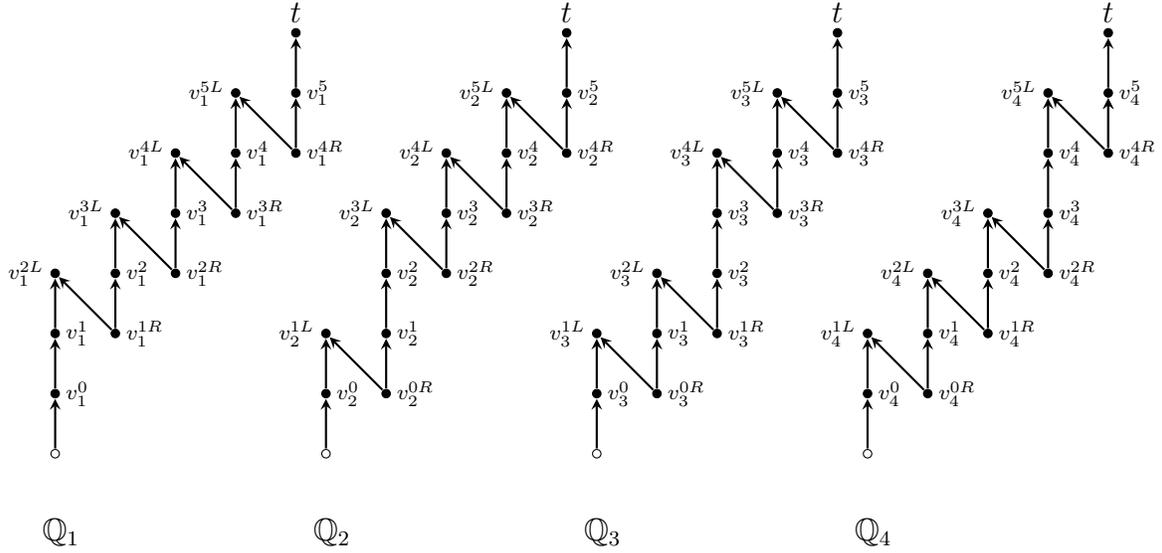
\begin{center}
\begin{tikzpicture}[scale=.8]

\def\x{-14}

\node at (15.1+\x,-1.3) {$\bbQ_1$};
\draw (15+\x,0) circle (2.0pt);
\draw [fill] (15+\x,1) circle (2.0pt);
\draw [fill] (15+\x,2) circle (2.0pt);
\draw [fill] (15+\x,3) circle (2.0pt);
\draw [fill] (16+\x,2) circle (2.0pt);
\draw [fill] (16+\x,3) circle (2.0pt);
\draw [fill] (16+\x,4) circle (2.0pt);
\draw [fill] (17+\x,3) circle (2.0pt);
\draw [fill] (17+\x,4) circle (2.0pt);
\draw [fill] (17+\x,5) circle (2.0pt);
\draw [fill] (18+\x,4) circle (2.0pt);
\draw [fill] (18+\x,5) circle (2.0pt);
\draw [fill] (18+\x,6) circle (2.0pt);
\draw [fill] (19+\x,5) circle (2.0pt);
\draw [fill] (19+\x,6) circle (2.0pt);
\draw [fill] (19+\x,7) circle (2.0pt);
\draw [->,>=stealth,thick,shorten >=2pt,shorten <=2pt] (15+\x,0) -- (15+\x,1);
\draw [->,>=stealth,thick,shorten >=2pt,shorten <=2pt] (15+\x,1) -- (15+\x,2);
\draw [->,>=stealth,thick,shorten >=2pt,shorten <=2pt] (15+\x,2) -- (15+\x,3);
\draw [->,>=stealth,thick,shorten >=2pt,shorten <=2pt] (16+\x,2) -- (15+\x,3);
\draw [->,>=stealth,thick,shorten >=2pt,shorten <=2pt] (16+\x,2) -- (16+\x,3);
\draw [->,>=stealth,thick,shorten >=2pt,shorten <=2pt] (16+\x,3) -- (16+\x,4);
\draw [->,>=stealth,thick,shorten >=2pt,shorten <=2pt] (17+\x,3) -- (16+\x,4);
\draw [->,>=stealth,thick,shorten >=2pt,shorten <=2pt] (17+\x,3) -- (17+\x,4);
\draw [->,>=stealth,thick,shorten >=2pt,shorten <=2pt] (17+\x,4) -- (17+\x,5);
\draw [->,>=stealth,thick,shorten >=2pt,shorten <=2pt] (18+\x,4) -- (17+\x,5);
\draw [->,>=stealth,thick,shorten >=2pt,shorten <=2pt] (18+\x,4) -- (18+\x,5);
\draw [->,>=stealth,thick,shorten >=2pt,shorten <=2pt] (18+\x,5) -- (18+\x,6);
\draw [->,>=stealth,thick,shorten >=2pt,shorten <=2pt] (19+\x,5) -- (18+\x,6);
\draw [->,>=stealth,thick,shorten >=2pt,shorten <=2pt] (19+\x,5) -- (19+\x,6);
\draw [->,>=stealth,thick,shorten >=2pt,shorten <=2pt] (19+\x,6) -- (19+\x,7);
\node [anchor=west] at (15+\x,1) {$\qlab 1 2a0$};
\node [anchor=west] at (15+\x,2) {$\qlab 1 2a1$};
\node [anchor=west] at (16+\x,2) {$\qlabR 1 2a1$};
\node [anchor=east] at (15+\x,3) {$\qlabL 1 2a2$};
\node [anchor=west] at (18+\x,4) {$\qlabR 1 2a3$};
\node [anchor=west] at (18+\x,5) {$\qlab 1 2a4$};
\node [anchor=east] at (18+\x,6) {$\qlabL 1 2a5$};
\node [anchor=east] at (17+\x,5) {$\qlabL 1 2a4$};
\node [anchor=west] at (19+\x,5) {$\qlabR 1 2a4$};
\node [anchor=west] at (19+\x,6) {$\qlab 1 2a5$};
\node [anchor=south] at (19+\x,7) {$\Qtop$};
\node [anchor=west] at (17+\x,4) {$\qlab 1 2a3$};
\node [anchor=east] at (16+\x,4) {$\qlabL 1 2a3$};
\node [anchor=west] at (16+\x,3) {$\qlab 1 2a2$};
\node [anchor=west] at (17+\x,3) {$\qlabR 1 2a2$};

\def\x{-9.5}
\node at (15.1+\x,-1.3) {$\bbQ_2$};
\draw (15+\x,0) circle (2.0pt);
\draw [fill] (16+\x,1) circle (2.0pt);
\draw [fill] (15+\x,1) circle (2.0pt);
\draw [fill] (15+\x,2) circle (2.0pt);
\draw [fill] (16+\x,2) circle (2.0pt);
\draw [fill] (16+\x,3) circle (2.0pt);
\draw [fill] (16+\x,4) circle (2.0pt);
\draw [fill] (17+\x,3) circle (2.0pt);
\draw [fill] (17+\x,4) circle (2.0pt);
\draw [fill] (17+\x,5) circle (2.0pt);
\draw [fill] (18+\x,4) circle (2.0pt);
\draw [fill] (18+\x,5) circle (2.0pt);
\draw [fill] (18+\x,6) circle (2.0pt);
\draw [fill] (19+\x,5) circle (2.0pt);
\draw [fill] (19+\x,6) circle (2.0pt);
\draw [fill] (19+\x,7) circle (2.0pt);
\draw [->,>=stealth,thick,shorten >=2pt,shorten <=2pt] (15+\x,0) -- (15+\x,1);
\draw [->,>=stealth,thick,shorten >=2pt,shorten <=2pt] (15+\x,1) -- (15+\x,2);
\draw [->,>=stealth,thick,shorten >=2pt,shorten <=2pt] (16+\x,1) -- (15+\x,2);
\draw [->,>=stealth,thick,shorten >=2pt,shorten <=2pt] (16+\x,1) -- (16+\x,2);
\draw [->,>=stealth,thick,shorten >=2pt,shorten <=2pt] (16+\x,2) -- (16+\x,3);
\draw [->,>=stealth,thick,shorten >=2pt,shorten <=2pt] (16+\x,3) -- (16+\x,4);
\draw [->,>=stealth,thick,shorten >=2pt,shorten <=2pt] (17+\x,3) -- (16+\x,4);
\draw [->,>=stealth,thick,shorten >=2pt,shorten <=2pt] (17+\x,3) -- (17+\x,4);
\draw [->,>=stealth,thick,shorten >=2pt,shorten <=2pt] (17+\x,4) -- (17+\x,5);
\draw [->,>=stealth,thick,shorten >=2pt,shorten <=2pt] (18+\x,4) -- (17+\x,5);
\draw [->,>=stealth,thick,shorten >=2pt,shorten <=2pt] (18+\x,4) -- (18+\x,5);
\draw [->,>=stealth,thick,shorten >=2pt,shorten <=2pt] (18+\x,5) -- (18+\x,6);
\draw [->,>=stealth,thick,shorten >=2pt,shorten <=2pt] (19+\x,5) -- (18+\x,6);
\draw [->,>=stealth,thick,shorten >=2pt,shorten <=2pt] (19+\x,5) -- (19+\x,6);
\draw [->,>=stealth,thick,shorten >=2pt,shorten <=2pt] (19+\x,6) -- (19+\x,7);
\node [anchor=west] at (15+\x,1) {$\qlab 2 2b0$};
\node [anchor=west] at (16+\x,2) {$\qlab 2 2b1$};
\node [anchor=west] at (18+\x,4) {$\qlabR 2 2b3$};
\node [anchor=west] at (19+\x,5) {$\qlabR 2 2b4$};
\node [anchor=west] at (18+\x,5) {$\qlab 2 2b4$};
\node [anchor=west] at (19+\x,6) {$\qlab 2 2b5$};
\node [anchor=east] at (17+\x,5) {$\qlabL 2 2b4$};
\node [anchor=east] at (18+\x,6) {$\qlabL 2 2b5$};
\node [anchor=south] at (19+\x,7) {$\Qtop$};
\node [anchor=east] at (15+\x,2) {$\qlabL 2 2b1$};
\node [anchor=west] at (16+\x,1) {$\qlabR 2 2b0$};
\node [anchor=west] at (17+\x,4) {$\qlab 2 2b3$};
\node [anchor=east] at (16+\x,4) {$\qlabL 2 2b3$};
\node [anchor=west] at (16+\x,3) {$\qlab 2 2b2$};
\node [anchor=west] at (17+\x,3) {$\qlabR 2 2b2$};

\def\x{-5}
\node at (15.1+\x,-1.3) {$\bbQ_3$};
\draw (15+\x,0) circle (2.0pt);
\draw [fill] (15+\x,1) circle (2.0pt);
\draw [fill] (15+\x,2) circle (2.0pt);
\draw [fill] (16+\x,1) circle (2.0pt);
\draw [fill] (16+\x,2) circle (2.0pt);
\draw [fill] (16+\x,3) circle (2.0pt);
\draw [fill] (17+\x,2) circle (2.0pt);
\draw [fill] (17+\x,3) circle (2.0pt);
\draw [fill] (17+\x,4) circle (2.0pt);
\draw [fill] (17+\x,5) circle (2.0pt);
\draw [fill] (18+\x,4) circle (2.0pt);
\draw [fill] (18+\x,5) circle (2.0pt);
\draw [fill] (18+\x,6) circle (2.0pt);
\draw [fill] (19+\x,5) circle (2.0pt);
\draw [fill] (19+\x,6) circle (2.0pt);
\draw [fill] (19+\x,7) circle (2.0pt);
\draw [->,>=stealth,thick,shorten >=2pt,shorten <=2pt] (15+\x,0) -- (15+\x,1);
\draw [->,>=stealth,thick,shorten >=2pt,shorten <=2pt] (15+\x,1) -- (15+\x,2);
\draw [->,>=stealth,thick,shorten >=2pt,shorten <=2pt] (16+\x,1) -- (15+\x,2);
\draw [->,>=stealth,thick,shorten >=2pt,shorten <=2pt] (16+\x,1) -- (16+\x,2);
\draw [->,>=stealth,thick,shorten >=2pt,shorten <=2pt] (16+\x,2) -- (16+\x,3);
\draw [->,>=stealth,thick,shorten >=2pt,shorten <=2pt] (17+\x,2) -- (16+\x,3);
\draw [->,>=stealth,thick,shorten >=2pt,shorten <=2pt] (17+\x,2) -- (17+\x,3);
\draw [->,>=stealth,thick,shorten >=2pt,shorten <=2pt] (17+\x,3) -- (17+\x,4);
\draw [->,>=stealth,thick,shorten >=2pt,shorten <=2pt] (17+\x,4) -- (17+\x,5);
\draw [->,>=stealth,thick,shorten >=2pt,shorten <=2pt] (18+\x,4) -- (17+\x,5);
\draw [->,>=stealth,thick,shorten >=2pt,shorten <=2pt] (18+\x,4) -- (18+\x,5);
\draw [->,>=stealth,thick,shorten >=2pt,shorten <=2pt] (18+\x,5) -- (18+\x,6);
\draw [->,>=stealth,thick,shorten >=2pt,shorten <=2pt] (19+\x,5) -- (18+\x,6);
\draw [->,>=stealth,thick,shorten >=2pt,shorten <=2pt] (19+\x,5) -- (19+\x,6);
\draw [->,>=stealth,thick,shorten >=2pt,shorten <=2pt] (19+\x,6) -- (19+\x,7);
\node [anchor=west] at (15+\x,1) {$\qlab 3 2c0$};
\node [anchor=west] at (16+\x,2) {$\qlab 3 2c1$};
\node [anchor=west] at (17+\x,2) {$\qlabR 3 2c1$};
\node [anchor=west] at (18+\x,4) {$\qlabR 3 2c3$};
\node [anchor=west] at (19+\x,5) {$\qlabR 3 2c4$};
\node [anchor=west] at (18+\x,5) {$\qlab 3 2c4$};
\node [anchor=west] at (19+\x,6) {$\qlab 3 2c5$};
\node [anchor=east] at (17+\x,5) {$\qlabL 3 2c4$};
\node [anchor=east] at (18+\x,6) {$\qlabL 3 2c5$};
\node [anchor=south] at (19+\x,7) {$\Qtop$};
\node [anchor=east] at (15+\x,2) {$\qlabL 3 2c1$};
\node [anchor=west] at (16+\x,1) {$\qlabR 3 2c0$};
\node [anchor=west] at (17+\x,4) {$\qlab 3 2c3$};
\node [anchor=east] at (16+\x,3) {$\qlabL 3 2c2$};
\node [anchor=west] at (17+\x,3) {$\qlab 3 2c2$};

\def\x{0.5}
\node at (14.1+\x,-1.3) {$\bbQ_4$};
\draw (14+\x,0) circle (2.0pt);
\draw [fill] (14+\x,1) circle (2.0pt);
\draw [fill] (14+\x,2) circle (2.0pt);
\draw [fill] (15+\x,1) circle (2.0pt);
\draw [fill] (15+\x,2) circle (2.0pt);
\draw [fill] (15+\x,3) circle (2.0pt);
\draw [fill] (16+\x,2) circle (2.0pt);
\draw [fill] (16+\x,3) circle (2.0pt);
\draw [fill] (16+\x,4) circle (2.0pt);
\draw [fill] (17+\x,3) circle (2.0pt);
\draw [fill] (17+\x,4) circle (2.0pt);
\draw [fill] (17+\x,5) circle (2.0pt);
\draw [fill] (18+\x,5) circle (2.0pt);
\draw [fill] (17+\x,6) circle (2.0pt);
\draw [fill] (18+\x,6) circle (2.0pt);
\draw [fill] (18+\x,7) circle (2.0pt);
\draw [->,>=stealth,thick,shorten >=2pt,shorten <=2pt] (14+\x,0) -- (14+\x,1);
\draw [->,>=stealth,thick,shorten >=2pt,shorten <=2pt] (14+\x,1) -- (14+\x,2);
\draw [->,>=stealth,thick,shorten >=2pt,shorten <=2pt] (15+\x,1) -- (14+\x,2);
\draw [->,>=stealth,thick,shorten >=2pt,shorten <=2pt] (15+\x,1) -- (15+\x,2);
\draw [->,>=stealth,thick,shorten >=2pt,shorten <=2pt] (15+\x,2) -- (15+\x,3);
\draw [->,>=stealth,thick,shorten >=2pt,shorten <=2pt] (16+\x,2) -- (15+\x,3);
\draw [->,>=stealth,thick,shorten >=2pt,shorten <=2pt] (16+\x,2) -- (16+\x,3);
\draw [->,>=stealth,thick,shorten >=2pt,shorten <=2pt] (16+\x,3) -- (16+\x,4);
\draw [->,>=stealth,thick,shorten >=2pt,shorten <=2pt] (17+\x,3) -- (16+\x,4);
\draw [->,>=stealth,thick,shorten >=2pt,shorten <=2pt] (17+\x,3) -- (17+\x,4);
\draw [->,>=stealth,thick,shorten >=2pt,shorten <=2pt] (17+\x,4) -- (17+\x,5);
\draw [->,>=stealth,thick,shorten >=2pt,shorten <=2pt] (17+\x,5) -- (17+\x,6);
\draw [->,>=stealth,thick,shorten >=2pt,shorten <=2pt] (18+\x,5) -- (17+\x,6);
\draw [->,>=stealth,thick,shorten >=2pt,shorten <=2pt] (18+\x,5) -- (18+\x,6);
\draw [->,>=stealth,thick,shorten >=2pt,shorten <=2pt] (18+\x,5) -- (18+\x,6);
\draw [->,>=stealth,thick,shorten >=2pt,shorten <=2pt] (18+\x,6) -- (18+\x,7);
\node [anchor=west] at (14+\x,1) {$\qlab 4 2d0$};
\node [anchor=west] at (15+\x,2) {$\qlab 4 2d1$};
\node [anchor=west] at (16+\x,2) {$\qlabR 4 2d1$};
\node [anchor=east] at (15+\x,3) {$\qlabL 4 2d2$};
\node [anchor=west] at (18+\x,5) {$\qlabR 4 2d4$};
\node [anchor=west] at (17+\x,5) {$\qlab 4 2d4$};
\node [anchor=west] at (18+\x,6) {$\qlab 4 2d5$};
\node [anchor=east] at (17+\x,6) {$\qlabL 4 2d5$};
\node [anchor=south] at (18+\x,7) {$\Qtop$};
\node [anchor=east] at (14+\x,2) {$\qlabL 4 2d1$};
\node [anchor=west] at (15+\x,1) {$\qlabR 4 2d0$};
\node [anchor=west] at (17+\x,4) {$\qlab 4 2d3$};
\node [anchor=east] at (16+\x,4) {$\qlabL 4 2d3$};
\node [anchor=west] at (16+\x,3) {$\qlab 4 2d2$};
\node [anchor=west] at (17+\x,3) {$\qlabR 4 2d2$};

\end{tikzpicture}
\end{center}
(There is also a $\bbQ_5$ completing the pattern, but we will not need it.)  If we define
the bijection $\varR\mapsto \barvarR$ from $\{\verta,\vertb,\vertc,\vertd,
\vertt\}$ to $R$ via
\begin{eqnarray*}
\barverta &=& (0,0,0,1,0)\\
\barvertb &=& (0,1,1,0,0)\\
\barvertc &=& (1,0,1,0,0)\\
\barvertd &=& (1,1,0,1,0)\\
\barvertt &=& (2,2,2,2,0)
\end{eqnarray*}
then the following is true (see \cite[proof of Lemma~3.6]{bdjn}).

\begin{clm} \label{clm1}
for all $\varA \in \{0,1,2\}$, $i \in \{1,2,3,4\}$
and $\varR \in \{\verta,\vertb,\vertc,\vertd,\vertt\}$,
\[
\bbQ_i \ra \bbP_{\varA,\varR} \quad\mbox{iff}\quad \barvarR[i]=\varA.
\]
Moreover, if $\barvarR[i]=\varA$ then the homomorphism 
$\bbQ_i\ra \bbP_{\varA,\varR}$ is unique
and surjective.  
\end{clm}

Next I define two gadgets: $\bbS(x,y,z)$ and $\bbT(x,y,z)$.  These are the digraphs
with three distinguished vertices $x,y,z$ given in the next figure:

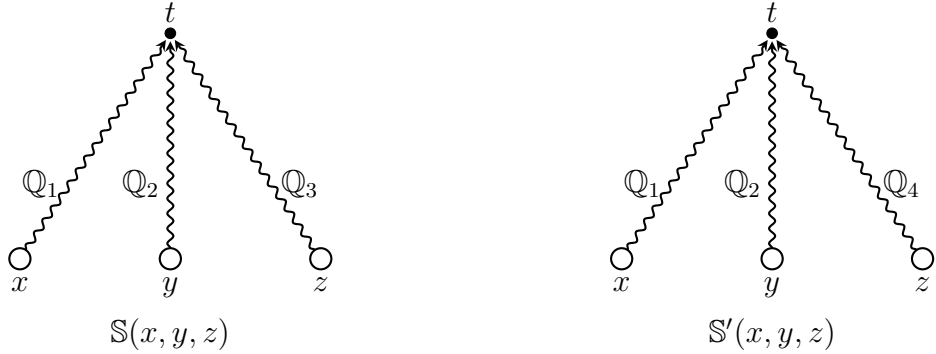
\begin{center}
\begin{tikzpicture}


\draw  [thick] (0,0) circle (4.0pt);
\draw  [thick] (2,0) circle (4.0pt);
\draw  [thick] (4,0) circle (4.0pt);

\draw  [thick] (8,0) circle (4.0pt);
\draw  [thick] (12,0) circle (4.0pt);
\draw  [thick] (10,0) circle (4.0pt);

\draw [fill] (2,3) circle (2.0pt);
\draw [fill] (10,3) circle (2.0pt);

\node [anchor=south] at (2,3) {$\Qtop$};
\node [anchor=south] at (10,3) {$\Qtop$};

\draw [->,line before snake=4pt,shorten <=4pt,snake=snake,segment amplitude=.4mm,segment length=2mm,line after snake=5pt,>=stealth,thick,shorten >=3pt] 
(0,0) -- (2,3);
\draw [->,line before snake=4pt,shorten <=4pt,snake=snake,segment amplitude=.4mm,segment length=2mm,line after snake=5pt,>=stealth,thick,shorten >=3pt] 
(2,0) -- (2,3);
\draw [->,line before snake=4pt,shorten <=4pt,snake=snake,segment amplitude=.4mm,segment length=2mm,line after snake=5pt,>=stealth,thick,shorten >=3pt] 
(4,0) -- (2,3);

\draw [->,line before snake=4pt,shorten <=4pt,snake=snake,segment amplitude=.4mm,segment length=2mm,line after snake=4pt,>=stealth,thick,shorten >=3pt] 
(8,0) -- (10,3);
\draw [->,line before snake=4pt,shorten <=4pt,snake=snake,segment amplitude=.4mm,segment length=2mm,line after snake=4pt,>=stealth,thick,shorten >=3pt] 
(10,0) -- (10,3);
\draw [->,line before snake=4pt,shorten <=4pt,snake=snake,segment amplitude=.4mm,segment length=2mm,line after snake=4pt,>=stealth,thick,shorten >=3pt] 
(12,0) -- (10,3);

\node [anchor=east] at (.67,1) {$\bbQ_1$};
\node [anchor=east] at (2,1) {$\bbQ_2$};
\node [anchor=west] at (3.33,1) {$\bbQ_3$};
\node [anchor=east] at (8.67,1) {$\bbQ_1$};
\node [anchor=east] at (10,1) {$\bbQ_2$};
\node [anchor=west] at (11.33,1) {$\bbQ_4$};

\node [anchor=north] at (0,-.1) {$x$};
\node [anchor=north] at (2,-.1) {$y$};
\node [anchor=north] at (4,-.1) {$z$};
\node [anchor=north] at (8,-.1) {$x$};
\node [anchor=north] at (10,-.1) {$y$};
\node [anchor=north] at (12,-.1) {$z$};

\node at (2,-1) {$\bbS(x,y,z)$};
\node at (10,-1) {$\bbT(x,y,z)$};
\end{tikzpicture}
\end{center}

The 3-ary relations on $V(\bbH)$ pp-defined by $\bbS(x,y,z)$ and $\bbT(x,y,z)$,
\begin{eqnarray*}
S &:=& \set{h(x),h(y),h(z))}{$\bbS(x,y,z)\stackrel{h}{\ra}\bbH$} \\
S' &:=& \set{h(x),h(y),h(z))}{$\bbT(x,y,z)\stackrel{h}{\ra}\bbH$}, 
\end{eqnarray*}
turn out to be
\begin{eqnarray*}
S &=& \proj_{1,2,3}(R) 
~=~ \set{(u,v,w) \in (\bbZ_2)^3}{$u+v+w=0$} \cup \{(2,2,2)\}\\
S' &=& \proj_{1,2,4}(R) 
~=~ \set{(u,v,w) \in (\bbZ_2)^3}{$u+v+w=1$} \cup \{(2,2,2)\}.
\end{eqnarray*}
Moreover, 
for each $(a,b,c) \in S$ there exists a unique homomorphim 
$\bbS(x,y,z)\stackrel{h}{\ra} \bbH$ satisfying $(h(x),h(y),h(z))=(a,b,c)$.
A similar
remark holds for $S'$ and $\bbT(x,y,z)$.


Now define $\bbG$ to be the digraph obtained by starting from vertices
$\{x_1,x_2,\ldots,x_6\}$ and connecting them with the following four gadgets:
\[
\bbS(x_1,x_2,x_3),\quad
\bbS(x_1,x_5,x_6),\quad
\bbS(x_4,x_2,x_6),\quad
\bbT(x_4,x_5,x_3).
\]
(Of course one takes pairwise disjoint isomorphic copies of the gadgets.)
Thus $\bbG$ is the digraph depicted schematically in the next figure:

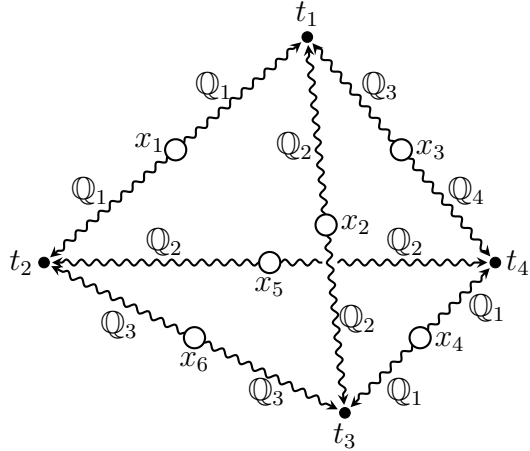
\begin{center}
\begin{tikzpicture}

\draw [fill] (3.5,3) circle (2.0pt); 
\draw [fill] (0,0) circle (2.0pt);   
\draw [fill] (4,-2) circle (2.0pt);  
\draw [fill] (6,0) circle (2.0pt);   

\draw [thick] (1.75,1.5) circle (4.0pt);   
\draw [thick] (3.75,0.5) circle (4.0pt);   
\draw [thick] (2,-1)     circle (4.0pt);   
\draw [thick] (4.75,1.5) circle (4.0pt);   
\draw [thick] (3,0)      circle (4.0pt);   
\draw [thick] (5,-1)     circle (4.0pt);   

\node [anchor=east] at (2.65,2.35) {$\bbQ_1$};
\node [anchor=east] at (3.725,1.55) {$\bbQ_2$};
\node [anchor=west] at (4.075,2.35) {$\bbQ_3$};

\node [anchor=south] at (1.6,-.05) {$\bbQ_2$};
\node [anchor=east] at (1.0,.95) {$\bbQ_1$};
\node [anchor=north] at (1,-.5) {$\bbQ_3$};

\node [anchor=north] at (2.96,-1.4) {$\bbQ_3$};
\node [anchor=north] at (4.8,-1.4) {$\bbQ_1$};
\node [anchor=west] at (3.775,-.75) {$\bbQ_2$};

\node [anchor=west] at (5.225,.95) {$\bbQ_4$};
\node [anchor=south] at (4.8,-.05) {$\bbQ_2$};
\node [anchor=west] at (5.5,-.6) {$\bbQ_1$};

\node [anchor=south] at (3.5,3) {$\Qtop_1$};
\node [anchor=east] at (0,0) {$\Qtop_2$};
\node [anchor=north] at (4,-2) {$\Qtop_3$};
\node [anchor=west] at (6,0) {$\Qtop_4$};

\node [anchor=east] at (1.75,1.6) {$x_1$};
\node [anchor=west] at (3.80,0.5) {$x_2$};
\node [anchor=north] at (2,-1.05) {$x_6$};
\node [anchor=west] at (4.80,1.5) {$x_3$};
\node [anchor=north] at (3,-.05) {$x_5$};
\node [anchor=west] at (5.05,-1.05) {$x_4$};

\draw [->,line before snake=4pt,shorten <=4pt,snake=snake,segment amplitude=.4mm,segment length=2mm,line after snake=5pt,>=stealth,thick,shorten >=3pt] 
(1.75,1.5) -- (3.5,3);

\draw [->,line before snake=4pt,shorten <=4pt,snake=snake,segment amplitude=.4mm,segment length=2mm,line after snake=5pt,>=stealth,thick,shorten >=3pt] 
(1.75,1.5) -- (0,0);

\draw [->,line before snake=4pt,shorten <=4pt,snake=snake,segment amplitude=.4mm,segment length=2mm,line after snake=5pt,>=stealth,thick,shorten >=3pt] 
(3.75,0.5) -- (3.5,3);

\draw [->,line before snake=4pt,shorten <=4pt,snake=snake,segment amplitude=.4mm,segment length=2mm,line after snake=5pt,>=stealth,thick,shorten >=3pt] 
(3.75,0.5) -- (4,-2);

\draw [->,line before snake=4pt,shorten <=4pt,snake=snake,segment amplitude=.4mm,segment length=2mm,line after snake=5pt,>=stealth,thick,shorten >=3pt] 
(2,-1) -- (0,0);

\draw [->,line before snake=4pt,shorten <=4pt,snake=snake,segment amplitude=.4mm,segment length=2mm,line after snake=5pt,>=stealth,thick,shorten >=3pt] 
(2,-1) -- (4,-2);

\draw [->,line before snake=4pt,shorten <=4pt,snake=snake,segment amplitude=.4mm,segment length=2mm,line after snake=5pt,>=stealth,thick,shorten >=3pt] 
(4.75,1.5) -- (3.5,3);

\draw [->,line before snake=4pt,shorten <=4pt,snake=snake,segment amplitude=.4mm,segment length=2mm,line after snake=5pt,>=stealth,thick,shorten >=3pt] 
(4.75,1.5) -- (6,0);

\draw [->,line before snake=4pt,shorten <=4pt,snake=snake,segment amplitude=.4mm,segment length=2mm,line after snake=5pt,>=stealth,thick,shorten >=3pt] 
(3,0) -- (0,0);

\draw [-,thick,line before snake=4pt,shorten <=4pt,snake=snake,segment amplitude=.4mm,segment length=2mm] 
(3,0) -- (3.7,0);

\draw [->,snake=snake,segment amplitude=.4mm,segment length=2mm,line after snake=5pt,>=stealth,thick,shorten >=3pt] 
(3.9,0) -- (6,0); 

\draw [->,line before snake=4pt,shorten <=4pt,snake=snake,segment amplitude=.4mm,segment length=2mm,line after snake=5pt,>=stealth,thick,shorten >=3pt] 
(5,-1) -- (6,0);

\draw [->,line before snake=4pt,shorten <=4pt,snake=snake,segment amplitude=.4mm,segment length=2mm,line after snake=5pt,>=stealth,thick,shorten >=3pt] 
(5,-1) -- (4,-2);

\end{tikzpicture}
\end{center}

Using Claim~\ref{clm1} and the definition of $\bbG$, one can show that
there is exactly one homomorphism $h:\bbG\ra \bbH$
and it satisfies $h(x_i)=2$ for all $i=1,2,\ldots,6$.

\subsection{The unary lists and $\phi'$} \label{list-sec}

In this section I describe the unary lists $L(x)$ which are produced by
enforcing (2,3)-consistency on $\bbG\stackrel{?}{\ra}\bbH$.
I also describe the restrictions $\phi'|_{L(x)}$ to these lists
of any of the idempotent cyclic polymorphisms
$\phi'$ of $\bbH$ produced by the recipe in \cite{bdjn}.

For each $\varA \in \{0,1,2\}$, $i \in \{1,2,3,4\}$, and $\varR \in
\{\verta,\vertb,\vertc,\vertd,\vertt\}$ with $\barvarR[i]=\varA$,
let $h_i^{\varA,\varR}$ denote the unique homomorphism
$\bbQ_i \ra \bbP_{\varA,\varR}$.  (See Claim~\ref{clm1}.)  

Let $B = \{\verta,\vertb,\vertc,\vertd,\vertt\}$.
For each $i=1,\ldots,4$ and $v \in V(\bbQ_i)$,
define $\sigma_{i,v}:B \ra V(\bbH)$ by
\[
\sigma_{i,v}(\varR) := h_i^{\varA,\varR}(v)\quad
\mbox{where $\varA=\barvarR[i]$},
\]
and note that $\sigma_{i,v}(\varR)$ is a vertex of the path 
$\bbP_{\varA,\lambda}$ from $a$ to $\varR$, and is at the same
height as $v$ is in $\bbQ_i$.  In particular, $\sigma_{i,\Qtop}$
is the identity map $B\ra B$ for all $i$.

Every vertex $\barv$ of $\bbG$ with $\barv \not\in \{x_1,\ldots,x_6\}$
is the image of some $v \in V(\bbQ_i)$ in
the copy of $\bbQ_i$ ending at $t_j$ for some unique $v$ and $j$.  
If in addition $\barv \not\in \{t_1,t_2,t_3,t_4\}$
(i.e., $v\ne t$), then $i$ is also uniquely determined.  In all cases
I'll let $\sigma_{\barv} := \sigma_{i,v}$.

\begin{clm} \label{clm-unary}
Let $L(v)$ be the unary lists produced by enforcing $(2,3)$-consistency
on $\bbG\stackrel{?}{\ra}\bbH$.
\begin{enumerate}
\item
$L(x_i) = \{0,1,2\}$ for all $i=1,\ldots,6$.
\item
For every
$\barv \in V(\bbG)\setminus \{x_1,\ldots,x_6\}$,
$L(\barv) = \ran(\sigma_{\barv})$.  
Moreover, $\sigma_{\barv}:B\ra L(\barv)$ is a bijection.
\end{enumerate}
\end{clm}

In particular, $L(t_j) = B$ for each $j=1,\ldots,4$.  As another
example, if $v = \bigqlabR 12 \in V(\bbQ_1)$ and $\barv$ is its image
in the copy of $\bbQ_1$ from $x_4$ to $t_3$ in $\bbG$, then 
\begin{eqnarray*}
L(\barv) ~=~ \ran(\sigma_{1,v}) &=&  \{\,
\sigma_{1,v}(\verta),~\sigma_{1,v}(\vertb),~\sigma_{1,v}(\vertc),~
\sigma_{1,v}(\vertd),~\sigma_{1,v}(\vertt)\,\}\\
&=& \{\, h_1^{0,\verta}(v),~h_1^{0,\vertb}(v),~h_1^{1,\vertc}(v),~
h_1^{1,\vertd}(v),~h_1^{2,\vertt}(v)\,\}\\
&=& \{\, \biglab 0{\verta}2,~~\biglabR0{\vertb}2,~~\biglab 1{\vertc}2,~~
\biglabR 1{\vertd}2,~~\biglab 2{\vertt}2\,\}.
\end{eqnarray*}

The recipe from \cite{bdjn} for producing a 3-ary idempotent cyclic polymorphism
$\phi'$ of $\bbH$ extending $\phi$ is not canonical; it depends on a choice
of such a polymorphism of the digraph 
$\bullet{\ra}\bullet{\leftarrow}\bullet{\ra}\bullet$.  Luckily, the restrictions
of $\phi'$ to the lists $L(v)$ are independent of this choice.

\begin{clm}
Let $\phi'$ be one of the $3$-ary idempotent cyclic polymorphisms of $\bbH$
extending $\phi$ given by the construction in \cite{bdjn}.
\begin{enumerate}
\item
$\phi'|_{\{0,1,2\}} = \phi$.
\item
The bijection $\varR \mapsto \barvarR$ from
$B$ to $R \subseteq A^5$ 
 is an isomorphism from the algebra $\langle B,\phi'|_B\rangle$
to the subalgebra of $\langle A,\phi\rangle^5$ with domain $R$.
\item
Suppose $\barv \in V(\bbG)\setminus \{x_1,\ldots,x_6\}$.  Then $\sigma_{\barv}$
is an isomorphism from $\langle B,\phi'_B\rangle$ to $\langle
L(\barv),\phi'_{L(\barv)}\rangle$.
\end{enumerate}
\end{clm}

At this point the reader has enough information to verify that $\bbH$,
$\bbG$, and $\phi'$ satisfy Claim~\ref{fkr-clm}.

\subsection{The binary lists}

For completeness, I describe the binary lists that arise from enforcing
(2,3)-consistency on $\bbG\stackrel{?}{\ra}\bbH$.
I first define some binary relations on $A$ ($=\{0,1,2\}$) and $B$.

\begin{df}
Here are some binary relations on $B$:
\begin{eqnarray*}
E_1 &=& 
\{\verta,\vertb\}^2 \cup \{\vertc,\vertd\}^2 \cup \{(\vertt,\vertt)\}\\
E_2 &=& 
\{\verta,\vertc\}^2 \cup \{\vertb,\vertd\}^2 \cup \{(\vertt,\vertt)\}\\
E_3 &=& 
\{\verta,\vertd\}^2 \cup \{\vertb,\vertc\}^2 \cup \{(\vertt,\vertt)\}\\
E_{34} &=&
(\{\verta,\vertd\}\times \{\vertb,\vertc\}) \cup 
(\{\vertb,\vertc\}\times \{\verta,\vertd\}) \cup  \{(\vertt,\vertt)\}.
\end{eqnarray*}
Here are some relations between $B$ and $A$ (subsets of $B\times A$):
\begin{eqnarray*}
P_1 &=& \{(\verta,0),(\vertb,0),(\vertc,1),(\vertd,1),(\vertt,2)\}\\
P_2 &=& \{(\verta,0),(\vertb,1),(\vertc,0),(\vertd,1),(\vertt,2)\}\\
P_3 &=& \{(\verta,0),(\vertb,1),(\vertc,1),(\vertd,0),(\vertt,2)\}\\
P_4 &=& \{(\verta,1),(\vertb,0),(\vertc,0),(\vertd,1),(\vertt,2)\}\\
\Delta_{BA} &=& (\{\verta,\vertb,\vertc,\vertd\} \times \{0,1\}) \cup
\{(\vertt,2)\}.
\end{eqnarray*}
Finally, I need the relation $\Delta = \{0,1\}^2 \cup \{(2,2)\}$ on $A$.
\end{df}

\begin{clm} \label{clm-binary}
Let $L(v,v')$ be the binary lists produced by 
enforcing $(2,3)$-consistency on $\bbG\stackrel{?}{\ra} \bbH$.
\begin{enumerate}
\item
$L(x_i,x_j) = \Delta$ for all $i\ne j$.  
\item
If $x_k$ and $t_j$ are the endpoints of a copy of $\bbQ_i$, then
$L(t_j,x_k) = P_i$.
\item
If $x_k$ and $t_j$ are not the endpoints of a copy of any $\bbQ_i$,
then $L(t_j,x_k) = \Delta_{BA}$.
\item
If $t_j$ and $t_k$ ($j\ne k$) are connected by two copies of $\bbQ_i$ extending
from some $x_\ell$, then $L(t_j,t_k) = E_i$.
In the remaining case, $L(t_1,t_4) = E_{34}$.
\item
Suppose $\barv \in V(\bbG) \setminus \{x_1,\ldots,x_6\}$ such that $\barv$
is in a copy of $\bbQ_i$ ending at $t_j$.  Then $L(t_j,\barv) = \graph(\sigma_{\barv})$.
\end{enumerate}
\end{clm}

All other values of $L(v,v')$ can be deduced from this claim.

\section{Discussion}

Using Claims~\ref{clm-unary} and \ref{clm-binary}, one can
see that each unary list $L(v)$ produced by the example in
Claim~\ref{fkr-clm} can be partitioned into
two sublists $L_{01}(v)$ and $L_2(v)$, so that
for all $v,v' \in V(\bbG)$, $L(v,v')$ is contained in 
$(L_{01}(v) \times L_{01}(v'))\cup (L_2(v) \times L_2(v'))$.
In other words, the microstructure graph of $(\bbG,\bbH,L)$ disconnects into
two components.
The components can be easily found, and the question $\bbG\stackrel{?}{\ra}
\bbH$ can thus be reduced to searching in the lists $L_{01}(v)$ and
(separately) in the lists $L_2(v)$.  $\phi'$ restricted to either 
family of sublists is already in the Mal'tsev case, so one can invoke
the Bulatov-Dalmau algorithm on each sublist to determine that $\bbG\ra\bbH$.
(An analogous fact is true of the original CSP($\bbA'$) instance defined in
Section~\ref{subsec-high}.)

Any reasonable algorithm can be assumed to first look for this kind of 
decomposition.  In particular, Feder, Kinne and Rafiey could insert a test 
for this kind of decomposition in their algorithm-idea; the instance 
constructed in Claim~\ref{fkr-clm} would then not refute their (modified)
algorithm.

In the next section I will sketch the construction of some 
more complicated examples which do not decompose in this trivial way.

\section{A more complicated example}

This second counter-example is a variation of the first, so I won't describe
it quite as thoroughly.  Here is the idea on which it is based.  Take the
template $\bbA=\langle A;E,R_0\rangle$ with domain
$A=\{0,1,2\}$ and two relations $E$ (2-ary) and $R_0$ (4-ary) given by
\begin{eqnarray*}
E &=& \{0,1,2\}^2 \setminus \{(2,2)\}\\
R_0 &=&\{(0,0,0,1),(0,1,1,0,(1,0,1,0, (1,1,0,1,(2,2,2,2)\}.
\end{eqnarray*}
$\bbA$ is core, and the operation
$\phi:A^3\ra A$ defined below is a WNU polymorphism:

\[
\phi(x,y,z) = \left\{\begin{array}{cl}
m(x,y,z) & \mbox{if $\{x,y,z\} \subseteq \{0,1\}$}\\
a & \mbox{if $(x,y,z) \in \{(2,a,b),(b,2,a),(a,b,2)\}$ with $a \in \{0,1\}$}\\
2 & \mbox{if $x=y=z=2$.}
\end{array}\right.
\]

Let $R_1$ and $R_2$ be the 3-ary relations obtained by projecting $R_0$ onto
coordinates $(1,2,3)$ and $(1,2,4)$ respectively.  
Let $\bbA'=\langle A,E,R_1,R_2\rangle$.
The CSP($\bbA'$) instance $\calI$
given by
\[
E(x_0,x_1) ~\&~ 
R_1(x_1,x_2,x_3) ~\&~ R_1(x_1,x_5,x_6) ~\&~
R_1(x_2,x_4,x_6) ~\&~ R_2(x_3,x_4,x_5)
\]
is (2,3)-minimal, the microstructure graph of its lists is connected, 
and it has exactly two solutions:
the maps which send 
$\{x_1,\ldots,x_6\}\mapsto 2$ and $x_0 \mapsto 0$ or $1$.
Thus one cannot delete 2 from the list for any of $x_1,\ldots,x_6$ without
losing all solutions, yet (from the perspective of $\phi$) the
appearance of 2 in any of the lists is a violation Mal'tsev property.
When this template and instance are translated to balanced digraphs via
Bul\'in, Deli\'c, Jackson and Niven \cite{bdjn}, the result 
is an instance of a digraph homomorphism problem whose lists do not decompose
in the trivial way, and which has a solution but step (4) of the algorithm-idea
of Feder, Kinne and Rafiey will fail with high probability (assuming that
the first list to be shrunk at step (4) is chosen randomly).

\subsection{The translation}

Define a balanced digraph $\bbH$ as follows.  
Let $R$ be the following 5-ary relation on $\{0,1,2\}$:
\begin{eqnarray*}
R &=& \{ 00010,~00011,~00012,~01100,~01101,~01102,~10100,\\
&& \phantom{\{}10101,~10102,~
11010,~11011,~11012,~22220,~22221\}.
\end{eqnarray*}
(Note that $\proj_{1,2,3,4}(R) = R_0$ and $\proj_{1,5}(R)=E$, and that
$\phi$ is a polymorphism of $R$.)

The vertex set $V(\bbH)$ consists
of the union of $A=\{0,1,2\}$, $R$, and 672 auxiliary vertices lying on 
42 pairwise disjoint oriented paths of net length 7 connecting each pair in
$A\times R$.  Given $a \in A$ and $\lambda \in R$, the oriented
path $\bbP_{a,\lambda}$ connecting $a$ to $\lambda$ is defined as follows:
\begin{itemize}
\item
$\bbP_{a,\lambda}$ starts with $a \ra u_{a\lambda}^0$ and ends with
$u_{a\lambda}^5\ra \lambda$.
\item
Intermediate vertices of $\bbP_{a,\lambda}$ are $u_{a\lambda}^i$ for 
$i=1,\ldots,4$.
\item
For each $i=1,\ldots,5$, the oriented path from $u_{a\lambda}^{i-1}$ to
$u_{a\lambda}^i$ is $u_{a\lambda}^{i-1}\ra u_{a\lambda}^i$ if 
$\lambda[i]=a$ and is $u_{a\lambda}^{i-1}\ra
u_{a\lambda}^{iL} \leftarrow u_{a\lambda}^{i{-}1\,R}\ra u_{a\lambda}^i$
otherwise.
\end{itemize}
See Figure~\ref{fig1} for a picture of $\bbP_{a,\lambda}$ when
$a=1$ and $\lambda = 01102$, and also a schematic diagram of $\bbH$.
$\bbH$ has a 3-ary WNU polymorphism $\phi'$ extending $\phi$, as shown
in \cite{bdjn}.

\begin{figure}
\begin{tikzpicture}[scale=.8]

w  (3,0) circle (2.0pt);
\draw [fill] (3,1) circle (2.0pt);
\draw [fill] (3,2) circle (2.0pt);
\draw [fill] (4,1) circle (2.0pt);
\draw [fill] (4,2) circle (2.0pt);
\draw [fill] (4,3) circle (2.0pt);
\draw [fill] (4,4) circle (2.0pt);
\draw [fill] (4,5) circle (2.0pt);
\draw [fill] (5,4) circle (2.0pt);
\draw [fill] (5,5) circle (2.0pt);
\draw [fill] (5,6) circle (2.0pt);
\draw [fill] (6,5) circle (2.0pt);
\draw [fill] (6,6) circle (2.0pt);
\draw  (6,7) circle (2.0pt);
\draw [->,>=stealth,thick,shorten >=2pt,shorten <=2pt] (3,0) -- (3,1);
\draw [->,>=stealth,thick,shorten >=2pt,shorten <=2pt] (3,1) -- (3,2);
\draw [->,>=stealth,thick,shorten >=2pt,shorten <=2pt] (4,1) -- (3,2);
\draw [->,>=stealth,thick,shorten >=2pt,shorten <=2pt] (4,1) -- (4,2);
\draw [->,>=stealth,thick,shorten >=2pt,shorten <=2pt] (4,2) -- (4,3);
\draw [->,>=stealth,thick,shorten >=2pt,shorten <=2pt] (4,3) -- (4,4);
\draw [->,>=stealth,thick,shorten >=2pt,shorten <=2pt] (4,4) -- (4,5);
\draw [->,>=stealth,thick,shorten >=2pt,shorten <=2pt] (5,4) -- (4,5);
\draw [->,>=stealth,thick,shorten >=2pt,shorten <=2pt] (5,4) -- (5,5);
\draw [->,>=stealth,thick,shorten >=2pt,shorten <=2pt] (5,5) -- (5,6);
\draw [->,>=stealth,thick,shorten >=2pt,shorten <=2pt] (6,5) -- (5,6);
\draw [->,>=stealth,thick,shorten >=2pt,shorten <=2pt] (6,5) -- (6,6);
\draw [->,>=stealth,thick,shorten >=2pt,shorten <=2pt] (6,6) -- (6,7);
\node [anchor=north] at (3,0) {$1$};
\node [anchor=north] at (4.5,-1) {$\bbP_{1,\lambda}$};

\node [anchor=west] at (3,1) {$\labl 1b0$};
\node [anchor=east] at (3,2) {$\labLl 1b1$};
\node [anchor=west] at (4,1) {$\labRl 1b0$};
\node [anchor=west] at (4,2) {$\labl 1b1$};
\node [anchor=west] at (4,3) {$\labl 1b2$};
\node [anchor=west] at (4,4) {$\labl 1b3$};
\node [anchor=west] at (5,5) {$\labl 1b4$};
\node [anchor=west] at (6,5) {$\labRl 1b4$};
\node [anchor=east] at (4,5) {$\labLl 1b4$};
\node [anchor=south] at (6,7) {$\lambda=01102$};
\node [anchor=west] at (5,4) {$\labRl 1b3$};
\node [anchor=west] at (6,6) {$\labl 1b5$};
\node [anchor=east] at (5,6) {$\labLl 1b5$};

\def\x{9}

\draw [fill] (1+\x,0) circle (2.0pt);
\draw [fill] (4+\x,0) circle (2.0pt);
\draw [fill] (7+\x,0) circle (2.0pt);

\draw [fill] (0+\x,3) circle (2.0pt);
\draw [fill] (2+\x,3) circle (2.0pt);
\draw [fill] (4+\x,3) circle (2.0pt);
\draw [fill] (8+\x,3) circle (2.0pt);

\node [anchor=north] at (1+\x,0) {$0$};
\node [anchor=north] at (4+\x,0) {$1$};
\node [anchor=north] at (7+\x,0) {$2$};
\node [anchor=south] at (0+\x,3) {$00010$};
\node [anchor=south] at (2+\x,3) {$00011$};
\node [anchor=south] at (4+\x,3) {$00012$};
\node [anchor=south] at (6+\x,3) {$\dots$};
\node [anchor=south] at (8+\x,3) {$22221$};

\draw [->,snake=snake,segment amplitude=.4mm,segment length=2mm,line after snake=4pt,>=stealth,thick,shorten >=2pt] 
(1+\x,0) -- (0+\x,3);
\draw [->,snake=snake,segment amplitude=.4mm,segment length=2mm,line after snake=4pt,>=stealth,thick,shorten >=2pt] 
(4+\x,0) -- (0+\x,3);
\draw [->,snake=snake,segment amplitude=.4mm,segment length=2mm,line after snake=4pt,>=stealth,thick,shorten >=2pt] 
(7+\x,0) -- (0+\x,3);
\draw [->,snake=snake,segment amplitude=.4mm,segment length=2mm,line after snake=4pt,>=stealth,thick,shorten >=2pt] 
(1+\x,0) -- (2+\x,3);
\draw [->,snake=snake,segment amplitude=.4mm,segment length=2mm,line after snake=4pt,>=stealth,thick,shorten >=2pt] 
(4+\x,0) -- (2+\x,3);
\draw [->,snake=snake,segment amplitude=.4mm,segment length=2mm,line after snake=4pt,>=stealth,thick,shorten >=2pt] 
(7+\x,0) -- (2+\x,3);
\draw [->,snake=snake,segment amplitude=.4mm,segment length=2mm,line after snake=4pt,>=stealth,thick,shorten >=2pt] 
(1+\x,0) -- (4+\x,3);
\draw [->,snake=snake,segment amplitude=.4mm,segment length=2mm,line after snake=4pt,>=stealth,thick,shorten >=2pt] 
(4+\x,0) -- (4+\x,3);
\draw [->,snake=snake,segment amplitude=.4mm,segment length=2mm,line after snake=4pt,>=stealth,thick,shorten >=2pt] 
(7+\x,0) -- (4+\x,3);
\draw [->,snake=snake,segment amplitude=.4mm,segment length=2mm,line after snake=4pt,>=stealth,thick,shorten >=2pt] 
(1+\x,0) -- (8+\x,3);
\draw [->,snake=snake,segment amplitude=.4mm,segment length=2mm,line after snake=4pt,>=stealth,thick,shorten >=2pt] 
(4+\x,0) -- (8+\x,3);
\draw [->,snake=snake,segment amplitude=.4mm,segment length=2mm,line after snake=4pt,>=stealth,thick,shorten >=2pt] 
(7+\x,0) -- (8+\x,3);

\node [anchor=north] at (4+\x,-1) {$\bbH$};

\end{tikzpicture}
\caption{} \label{fig1}
\end{figure}

For each $j=1,\ldots,5$, let $\bbQ_j$ be the oriented path of net length 7
defined as follows: 
\begin{itemize}
\item
$\bbQ_i$ starts as $b \ra v_i^0$ and ends as $v_i^5\ra t$.
\item
Intermediate vertices of $\bbQ_i$ are $v_i^j$ for 
$j=1,\ldots,4$.
\item
For each $j=1,\ldots,5$, the oriented path from $v_i^{j-1}$ to
$v_i^j$ is $v_i^{j-1}\ra v_i^j$ if 
$i=j$ and is $v_i^{j-1}\ra
v_i^{jL} \leftarrow v_i^{j{-}1\,R}\ra v_i^j$
otherwise.
\end{itemize}
Four of these paths are pictured in Figure 2.  As in the first
counter-example, we have $\bbQ_i \ra \bbP_{a,\lambda}$ iff $\lambda[i]=a$,
and when both conditions hold, the homomorphism is unique and surjective.

\begin{figure}
\begin{tikzpicture}[scale=.8]

\def\x{-14}

\node at (15.1+\x,-1.3) {$\bbQ_1$};
\node [anchor=north] at (15+\x,0) {$b$};
\draw [fill] (15+\x,0) circle (2.0pt);
\draw [fill] (15+\x,1) circle (2.0pt);
\draw [fill] (15+\x,2) circle (2.0pt);
\draw [fill] (15+\x,3) circle (2.0pt);
\draw [fill] (16+\x,2) circle (2.0pt);
\draw [fill] (16+\x,3) circle (2.0pt);
\draw [fill] (16+\x,4) circle (2.0pt);
\draw [fill] (17+\x,3) circle (2.0pt);
\draw [fill] (17+\x,4) circle (2.0pt);
\draw [fill] (17+\x,5) circle (2.0pt);
\draw [fill] (18+\x,4) circle (2.0pt);
\draw [fill] (18+\x,5) circle (2.0pt);
\draw [fill] (18+\x,6) circle (2.0pt);
\draw [fill] (19+\x,5) circle (2.0pt);
\draw [fill] (19+\x,6) circle (2.0pt);
\draw [fill] (19+\x,7) circle (2.0pt);
\draw [->,>=stealth,thick,shorten >=2pt,shorten <=2pt] (15+\x,0) -- (15+\x,1);
\draw [->,>=stealth,thick,shorten >=2pt,shorten <=2pt] (15+\x,1) -- (15+\x,2);
\draw [->,>=stealth,thick,shorten >=2pt,shorten <=2pt] (15+\x,2) -- (15+\x,3);
\draw [->,>=stealth,thick,shorten >=2pt,shorten <=2pt] (16+\x,2) -- (15+\x,3);
\draw [->,>=stealth,thick,shorten >=2pt,shorten <=2pt] (16+\x,2) -- (16+\x,3);
\draw [->,>=stealth,thick,shorten >=2pt,shorten <=2pt] (16+\x,3) -- (16+\x,4);
\draw [->,>=stealth,thick,shorten >=2pt,shorten <=2pt] (17+\x,3) -- (16+\x,4);
\draw [->,>=stealth,thick,shorten >=2pt,shorten <=2pt] (17+\x,3) -- (17+\x,4);
\draw [->,>=stealth,thick,shorten >=2pt,shorten <=2pt] (17+\x,4) -- (17+\x,5);
\draw [->,>=stealth,thick,shorten >=2pt,shorten <=2pt] (18+\x,4) -- (17+\x,5);
\draw [->,>=stealth,thick,shorten >=2pt,shorten <=2pt] (18+\x,4) -- (18+\x,5);
\draw [->,>=stealth,thick,shorten >=2pt,shorten <=2pt] (18+\x,5) -- (18+\x,6);
\draw [->,>=stealth,thick,shorten >=2pt,shorten <=2pt] (19+\x,5) -- (18+\x,6);
\draw [->,>=stealth,thick,shorten >=2pt,shorten <=2pt] (19+\x,5) -- (19+\x,6);
\draw [->,>=stealth,thick,shorten >=2pt,shorten <=2pt] (19+\x,6) -- (19+\x,7);
\node [anchor=west] at (15+\x,1) {$\qlab 1 2a0$};
\node [anchor=west] at (15+\x,2) {$\qlab 1 2a1$};
\node [anchor=west] at (16+\x,2) {$\qlabR 1 2a1$};
\node [anchor=east] at (15+\x,3) {$\qlabL 1 2a2$};
\node [anchor=west] at (18+\x,4) {$\qlabR 1 2a3$};
\node [anchor=west] at (18+\x,5) {$\qlab 1 2a4$};
\node [anchor=east] at (18+\x,6) {$\qlabL 1 2a5$};
\node [anchor=east] at (17+\x,5) {$\qlabL 1 2a4$};
\node [anchor=west] at (19+\x,5) {$\qlabR 1 2a4$};
\node [anchor=west] at (19+\x,6) {$\qlab 1 2a5$};
\node [anchor=south] at (19+\x,7) {$\Qtop$};
\node [anchor=west] at (17+\x,4) {$\qlab 1 2a3$};
\node [anchor=east] at (16+\x,4) {$\qlabL 1 2a3$};
\node [anchor=west] at (16+\x,3) {$\qlab 1 2a2$};
\node [anchor=west] at (17+\x,3) {$\qlabR 1 2a2$};

\def\x{-9.5}
\node at (15.1+\x,-1.3) {$\bbQ_2$};
\node [anchor=north] at (15+\x,0) {$b$};
\draw [fill] (15+\x,0) circle (2.0pt);
\draw [fill] (16+\x,1) circle (2.0pt);
\draw [fill] (15+\x,1) circle (2.0pt);
\draw [fill] (15+\x,2) circle (2.0pt);
\draw [fill] (16+\x,2) circle (2.0pt);
\draw [fill] (16+\x,3) circle (2.0pt);
\draw [fill] (16+\x,4) circle (2.0pt);
\draw [fill] (17+\x,3) circle (2.0pt);
\draw [fill] (17+\x,4) circle (2.0pt);
\draw [fill] (17+\x,5) circle (2.0pt);
\draw [fill] (18+\x,4) circle (2.0pt);
\draw [fill] (18+\x,5) circle (2.0pt);
\draw [fill] (18+\x,6) circle (2.0pt);
\draw [fill] (19+\x,5) circle (2.0pt);
\draw [fill] (19+\x,6) circle (2.0pt);
\draw [fill] (19+\x,7) circle (2.0pt);
\draw [->,>=stealth,thick,shorten >=2pt,shorten <=2pt] (15+\x,0) -- (15+\x,1);
\draw [->,>=stealth,thick,shorten >=2pt,shorten <=2pt] (15+\x,1) -- (15+\x,2);
\draw [->,>=stealth,thick,shorten >=2pt,shorten <=2pt] (16+\x,1) -- (15+\x,2);
\draw [->,>=stealth,thick,shorten >=2pt,shorten <=2pt] (16+\x,1) -- (16+\x,2);
\draw [->,>=stealth,thick,shorten >=2pt,shorten <=2pt] (16+\x,2) -- (16+\x,3);
\draw [->,>=stealth,thick,shorten >=2pt,shorten <=2pt] (16+\x,3) -- (16+\x,4);
\draw [->,>=stealth,thick,shorten >=2pt,shorten <=2pt] (17+\x,3) -- (16+\x,4);
\draw [->,>=stealth,thick,shorten >=2pt,shorten <=2pt] (17+\x,3) -- (17+\x,4);
\draw [->,>=stealth,thick,shorten >=2pt,shorten <=2pt] (17+\x,4) -- (17+\x,5);
\draw [->,>=stealth,thick,shorten >=2pt,shorten <=2pt] (18+\x,4) -- (17+\x,5);
\draw [->,>=stealth,thick,shorten >=2pt,shorten <=2pt] (18+\x,4) -- (18+\x,5);
\draw [->,>=stealth,thick,shorten >=2pt,shorten <=2pt] (18+\x,5) -- (18+\x,6);
\draw [->,>=stealth,thick,shorten >=2pt,shorten <=2pt] (19+\x,5) -- (18+\x,6);
\draw [->,>=stealth,thick,shorten >=2pt,shorten <=2pt] (19+\x,5) -- (19+\x,6);
\draw [->,>=stealth,thick,shorten >=2pt,shorten <=2pt] (19+\x,6) -- (19+\x,7);
\node [anchor=west] at (15+\x,1) {$\qlab 2 2b0$};
\node [anchor=west] at (16+\x,2) {$\qlab 2 2b1$};
\node [anchor=west] at (18+\x,4) {$\qlabR 2 2b3$};
\node [anchor=west] at (19+\x,5) {$\qlabR 2 2b4$};
\node [anchor=west] at (18+\x,5) {$\qlab 2 2b4$};
\node [anchor=west] at (19+\x,6) {$\qlab 2 2b5$};
\node [anchor=east] at (17+\x,5) {$\qlabL 2 2b4$};
\node [anchor=east] at (18+\x,6) {$\qlabL 2 2b5$};
\node [anchor=south] at (19+\x,7) {$\Qtop$};
\node [anchor=east] at (15+\x,2) {$\qlabL 2 2b1$};
\node [anchor=west] at (16+\x,1) {$\qlabR 2 2b0$};
\node [anchor=west] at (17+\x,4) {$\qlab 2 2b3$};
\node [anchor=east] at (16+\x,4) {$\qlabL 2 2b3$};
\node [anchor=west] at (16+\x,3) {$\qlab 2 2b2$};
\node [anchor=west] at (17+\x,3) {$\qlabR 2 2b2$};

\def\x{-5}
\node at (15.1+\x,-1.3) {$\bbQ_3$};
\node [anchor=north] at (15+\x,0) {$b$};
\draw [fill] (15+\x,0) circle (2.0pt);
\draw [fill] (15+\x,1) circle (2.0pt);
\draw [fill] (15+\x,2) circle (2.0pt);
\draw [fill] (16+\x,1) circle (2.0pt);
\draw [fill] (16+\x,2) circle (2.0pt);
\draw [fill] (16+\x,3) circle (2.0pt);
\draw [fill] (17+\x,2) circle (2.0pt);
\draw [fill] (17+\x,3) circle (2.0pt);
\draw [fill] (17+\x,4) circle (2.0pt);
\draw [fill] (17+\x,5) circle (2.0pt);
\draw [fill] (18+\x,4) circle (2.0pt);
\draw [fill] (18+\x,5) circle (2.0pt);
\draw [fill] (18+\x,6) circle (2.0pt);
\draw [fill] (19+\x,5) circle (2.0pt);
\draw [fill] (19+\x,6) circle (2.0pt);
\draw [fill] (19+\x,7) circle (2.0pt);
\draw [->,>=stealth,thick,shorten >=2pt,shorten <=2pt] (15+\x,0) -- (15+\x,1);
\draw [->,>=stealth,thick,shorten >=2pt,shorten <=2pt] (15+\x,1) -- (15+\x,2);
\draw [->,>=stealth,thick,shorten >=2pt,shorten <=2pt] (16+\x,1) -- (15+\x,2);
\draw [->,>=stealth,thick,shorten >=2pt,shorten <=2pt] (16+\x,1) -- (16+\x,2);
\draw [->,>=stealth,thick,shorten >=2pt,shorten <=2pt] (16+\x,2) -- (16+\x,3);
\draw [->,>=stealth,thick,shorten >=2pt,shorten <=2pt] (17+\x,2) -- (16+\x,3);
\draw [->,>=stealth,thick,shorten >=2pt,shorten <=2pt] (17+\x,2) -- (17+\x,3);
\draw [->,>=stealth,thick,shorten >=2pt,shorten <=2pt] (17+\x,3) -- (17+\x,4);
\draw [->,>=stealth,thick,shorten >=2pt,shorten <=2pt] (17+\x,4) -- (17+\x,5);
\draw [->,>=stealth,thick,shorten >=2pt,shorten <=2pt] (18+\x,4) -- (17+\x,5);
\draw [->,>=stealth,thick,shorten >=2pt,shorten <=2pt] (18+\x,4) -- (18+\x,5);
\draw [->,>=stealth,thick,shorten >=2pt,shorten <=2pt] (18+\x,5) -- (18+\x,6);
\draw [->,>=stealth,thick,shorten >=2pt,shorten <=2pt] (19+\x,5) -- (18+\x,6);
\draw [->,>=stealth,thick,shorten >=2pt,shorten <=2pt] (19+\x,5) -- (19+\x,6);
\draw [->,>=stealth,thick,shorten >=2pt,shorten <=2pt] (19+\x,6) -- (19+\x,7);
\node [anchor=west] at (15+\x,1) {$\qlab 3 2c0$};
\node [anchor=west] at (16+\x,2) {$\qlab 3 2c1$};
\node [anchor=west] at (17+\x,2) {$\qlabR 3 2c1$};
\node [anchor=west] at (18+\x,4) {$\qlabR 3 2c3$};
\node [anchor=west] at (19+\x,5) {$\qlabR 3 2c4$};
\node [anchor=west] at (18+\x,5) {$\qlab 3 2c4$};
\node [anchor=west] at (19+\x,6) {$\qlab 3 2c5$};
\node [anchor=east] at (17+\x,5) {$\qlabL 3 2c4$};
\node [anchor=east] at (18+\x,6) {$\qlabL 3 2c5$};
\node [anchor=south] at (19+\x,7) {$\Qtop$};
\node [anchor=east] at (15+\x,2) {$\qlabL 3 2c1$};
\node [anchor=west] at (16+\x,1) {$\qlabR 3 2c0$};
\node [anchor=west] at (17+\x,4) {$\qlab 3 2c3$};
\node [anchor=east] at (16+\x,3) {$\qlabL 3 2c2$};
\node [anchor=west] at (17+\x,3) {$\qlab 3 2c2$};

\def\x{0.5}
\node at (14.1+\x,-1.3) {$\bbQ_4$};
\node [anchor=north] at (14+\x,0) {$b$};
\draw [fill] (14+\x,0) circle (2.0pt);
\draw [fill] (14+\x,1) circle (2.0pt);
\draw [fill] (14+\x,2) circle (2.0pt);
\draw [fill] (15+\x,1) circle (2.0pt);
\draw [fill] (15+\x,2) circle (2.0pt);
\draw [fill] (15+\x,3) circle (2.0pt);
\draw [fill] (16+\x,2) circle (2.0pt);
\draw [fill] (16+\x,3) circle (2.0pt);
\draw [fill] (16+\x,4) circle (2.0pt);
\draw [fill] (17+\x,3) circle (2.0pt);
\draw [fill] (17+\x,4) circle (2.0pt);
\draw [fill] (17+\x,5) circle (2.0pt);
\draw [fill] (18+\x,5) circle (2.0pt);
\draw [fill] (17+\x,6) circle (2.0pt);
\draw [fill] (18+\x,6) circle (2.0pt);
\draw [fill] (18+\x,7) circle (2.0pt);
\draw [->,>=stealth,thick,shorten >=2pt,shorten <=2pt] (14+\x,0) -- (14+\x,1);
\draw [->,>=stealth,thick,shorten >=2pt,shorten <=2pt] (14+\x,1) -- (14+\x,2);
\draw [->,>=stealth,thick,shorten >=2pt,shorten <=2pt] (15+\x,1) -- (14+\x,2);
\draw [->,>=stealth,thick,shorten >=2pt,shorten <=2pt] (15+\x,1) -- (15+\x,2);
\draw [->,>=stealth,thick,shorten >=2pt,shorten <=2pt] (15+\x,2) -- (15+\x,3);
\draw [->,>=stealth,thick,shorten >=2pt,shorten <=2pt] (16+\x,2) -- (15+\x,3);
\draw [->,>=stealth,thick,shorten >=2pt,shorten <=2pt] (16+\x,2) -- (16+\x,3);
\draw [->,>=stealth,thick,shorten >=2pt,shorten <=2pt] (16+\x,3) -- (16+\x,4);
\draw [->,>=stealth,thick,shorten >=2pt,shorten <=2pt] (17+\x,3) -- (16+\x,4);
\draw [->,>=stealth,thick,shorten >=2pt,shorten <=2pt] (17+\x,3) -- (17+\x,4);
\draw [->,>=stealth,thick,shorten >=2pt,shorten <=2pt] (17+\x,4) -- (17+\x,5);
\draw [->,>=stealth,thick,shorten >=2pt,shorten <=2pt] (17+\x,5) -- (17+\x,6);
\draw [->,>=stealth,thick,shorten >=2pt,shorten <=2pt] (18+\x,5) -- (17+\x,6);
\draw [->,>=stealth,thick,shorten >=2pt,shorten <=2pt] (18+\x,5) -- (18+\x,6);
\draw [->,>=stealth,thick,shorten >=2pt,shorten <=2pt] (18+\x,5) -- (18+\x,6);
\draw [->,>=stealth,thick,shorten >=2pt,shorten <=2pt] (18+\x,6) -- (18+\x,7);
\node [anchor=west] at (14+\x,1) {$\qlab 4 2d0$};
\node [anchor=west] at (15+\x,2) {$\qlab 4 2d1$};
\node [anchor=west] at (16+\x,2) {$\qlabR 4 2d1$};
\node [anchor=east] at (15+\x,3) {$\qlabL 4 2d2$};
\node [anchor=west] at (18+\x,5) {$\qlabR 4 2d4$};
\node [anchor=west] at (17+\x,5) {$\qlab 4 2d4$};
\node [anchor=west] at (18+\x,6) {$\qlab 4 2d5$};
\node [anchor=east] at (17+\x,6) {$\qlabL 4 2d5$};
\node [anchor=south] at (18+\x,7) {$\Qtop$};
\node [anchor=east] at (14+\x,2) {$\qlabL 4 2d1$};
\node [anchor=west] at (15+\x,1) {$\qlabR 4 2d0$};
\node [anchor=west] at (17+\x,4) {$\qlab 4 2d3$};
\node [anchor=east] at (16+\x,4) {$\qlabL 4 2d3$};
\node [anchor=west] at (16+\x,3) {$\qlab 4 2d2$};
\node [anchor=west] at (17+\x,3) {$\qlabR 4 2d2$};

\end{tikzpicture}
\caption{} \label{fig2}
\end{figure}

The gadgets $\bbS(x,y,z)$ and $\bbS'(x,y,z)$ are defined exactly as before.
A new gadget $\bbE(x,y)$ is defined by the scheme $x\stackrel{\bbQ_5}{\ra} t
\stackrel{\bbQ_1}{\leftarrow} y$.

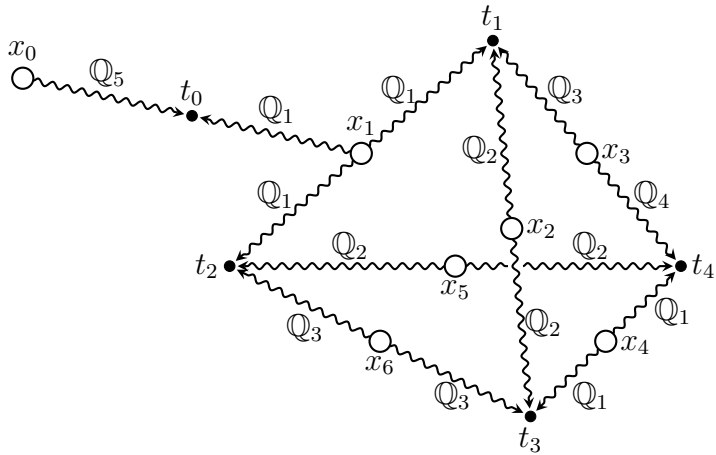
\begin{figure}
\begin{tikzpicture}

\draw [fill] (3.5,3) circle (2.0pt); 
\draw [fill] (0,0) circle (2.0pt);   
\draw [fill] (4,-2) circle (2.0pt);  
\draw [fill] (6,0) circle (2.0pt);   
\draw [fill] (-.5,2) circle (2.0pt);   

\draw [thick] (-2.75,2.5) circle (4.0pt);   
\draw [thick] (1.75,1.5) circle (4.0pt);   
\draw [thick] (3.75,0.5) circle (4.0pt);   
\draw [thick] (2,-1)     circle (4.0pt);   
\draw [thick] (4.75,1.5) circle (4.0pt);   
\draw [thick] (3,0)      circle (4.0pt);   
\draw [thick] (5,-1)     circle (4.0pt);   

\node [anchor=east] at (2.65,2.35) {$\bbQ_1$};
\node [anchor=east] at (3.725,1.55) {$\bbQ_2$};
\node [anchor=west] at (4.075,2.35) {$\bbQ_3$};

\node [anchor=south] at (1.6,-.05) {$\bbQ_2$};
\node [anchor=east] at (1.0,.95) {$\bbQ_1$};
\node [anchor=north] at (1,-.5) {$\bbQ_3$};

\node [anchor=north] at (2.96,-1.4) {$\bbQ_3$};
\node [anchor=north] at (4.8,-1.4) {$\bbQ_1$};
\node [anchor=west] at (3.775,-.75) {$\bbQ_2$};

\node [anchor=west] at (5.225,.95) {$\bbQ_4$};
\node [anchor=south] at (4.8,-.05) {$\bbQ_2$};
\node [anchor=west] at (5.5,-.6) {$\bbQ_1$};

\node [anchor=south] at (.625,1.75) {$\bbQ_1$};
\node [anchor=south] at (-1.625,2.25) {$\bbQ_5$};

\node [anchor=south] at (3.5,3) {$\Qtop_1$};
\node [anchor=south] at (-.5,2) {$\Qtop_0$};
\node [anchor=east] at (0,0) {$\Qtop_2$};
\node [anchor=north] at (4,-2) {$\Qtop_3$};
\node [anchor=west] at (6,0) {$\Qtop_4$};

\node [anchor=south] at (1.75,1.6) {$x_1$};
\node [anchor=south] at (-2.75,2.6) {$x_0$};
\node [anchor=west] at (3.80,0.5) {$x_2$};
\node [anchor=north] at (2,-1.05) {$x_6$};
\node [anchor=west] at (4.80,1.5) {$x_3$};
\node [anchor=north] at (3,-.05) {$x_5$};
\node [anchor=west] at (5.05,-1.05) {$x_4$};

\draw [->,line before snake=4pt,shorten <=4pt,snake=snake,segment amplitude=.4mm,segment length=2mm,line after snake=5pt,>=stealth,thick,shorten >=3pt] 
(1.75,1.5) -- (-.5,2);

\draw [->,line before snake=4pt,shorten <=4pt,snake=snake,segment amplitude=.4mm,segment length=2mm,line after snake=5pt,>=stealth,thick,shorten >=3pt] 
(-2.75,2.5) -- (-.5,2);

\draw [->,line before snake=4pt,shorten <=4pt,snake=snake,segment amplitude=.4mm,segment length=2mm,line after snake=5pt,>=stealth,thick,shorten >=3pt] 
(1.75,1.5) -- (3.5,3);

\draw [->,line before snake=4pt,shorten <=4pt,snake=snake,segment amplitude=.4mm,segment length=2mm,line after snake=5pt,>=stealth,thick,shorten >=3pt] 
(1.75,1.5) -- (0,0);

\draw [->,line before snake=4pt,shorten <=4pt,snake=snake,segment amplitude=.4mm,segment length=2mm,line after snake=5pt,>=stealth,thick,shorten >=3pt] 
(3.75,0.5) -- (3.5,3);

\draw [->,line before snake=4pt,shorten <=4pt,snake=snake,segment amplitude=.4mm,segment length=2mm,line after snake=5pt,>=stealth,thick,shorten >=3pt] 
(3.75,0.5) -- (4,-2);

\draw [->,line before snake=4pt,shorten <=4pt,snake=snake,segment amplitude=.4mm,segment length=2mm,line after snake=5pt,>=stealth,thick,shorten >=3pt] 
(2,-1) -- (0,0);

\draw [->,line before snake=4pt,shorten <=4pt,snake=snake,segment amplitude=.4mm,segment length=2mm,line after snake=5pt,>=stealth,thick,shorten >=3pt] 
(2,-1) -- (4,-2);

\draw [->,line before snake=4pt,shorten <=4pt,snake=snake,segment amplitude=.4mm,segment length=2mm,line after snake=5pt,>=stealth,thick,shorten >=3pt] 
(4.75,1.5) -- (3.5,3);

\draw [->,line before snake=4pt,shorten <=4pt,snake=snake,segment amplitude=.4mm,segment length=2mm,line after snake=5pt,>=stealth,thick,shorten >=3pt] 
(4.75,1.5) -- (6,0);

\draw [->,line before snake=4pt,shorten <=4pt,snake=snake,segment amplitude=.4mm,segment length=2mm,line after snake=5pt,>=stealth,thick,shorten >=3pt] 
(3,0) -- (0,0);

\draw [-,thick,line before snake=4pt,shorten <=4pt,snake=snake,segment amplitude=.4mm,segment length=2mm] 
(3,0) -- (3.7,0);

\draw [->,snake=snake,segment amplitude=.4mm,segment length=2mm,line after snake=5pt,>=stealth,thick,shorten >=3pt] 
(3.9,0) -- (6,0); 

\draw [->,line before snake=4pt,shorten <=4pt,snake=snake,segment amplitude=.4mm,segment length=2mm,line after snake=5pt,>=stealth,thick,shorten >=3pt] 
(5,-1) -- (6,0);

\draw [->,line before snake=4pt,shorten <=4pt,snake=snake,segment amplitude=.4mm,segment length=2mm,line after snake=5pt,>=stealth,thick,shorten >=3pt] 
(5,-1) -- (4,-2);

\end{tikzpicture}
\caption{The instance digraph $\bbG$} \label{fig3}
\end{figure}

Now define $\bbG$ to be the digraph obtained by starting from vertices
$\{x_1,x_1,\ldots,x_6\}$ and connecting them with the following five gadgets:
\[
\bbE(x_0,x_1),\quad
\bbS(x_1,x_2,x_3),\quad
\bbS(x_1,x_5,x_6),\quad
\bbS(x_4,x_2,x_6),\quad
\bbT(x_4,x_5,x_3).
\]
(Of course one takes pairwise disjoint isomorphic copies of the gadgets.)
Thus $\bbG$ is the digraph depicted schematically in Figure~\ref{fig3}.

By construction, 
there exist exactly two homomorphisms $\bbG\ra \bbH$, which are uniquely 
determined by their values on $x_0,\ldots,x_6$.  These homomorphisms 
correspond to the two solutions to the CSP($\bbA'$) instance
$\calI$ defined at the beginning of this section.  
Furthermore, when the instance $\bbG\stackrel{?}{\ra}\bbH$
is preprocessed, the microstructure graph of the resulting unary and
binary lists is highly connected.

This example is problematic for the Feder-Kinne-Rafiey algorithm-idea
for the following reason.  The two solutions agree
on the vertices $x_1,\ldots,x_6$ as well as on the paths connecting $x_1,
\ldots,x_6$ to $t_1,\ldots,t_4$.  The common value of the two solutions
at each of these vertices is also the unique element of the corresponding list
which fails the Mal'tsev property.  Thus if the Feder-Kinne-Rafiey
algorithm-idea is executed on this instance, and the first list chosen by
step (4) to be shrunk is \emph{not} on the path from $x_1$ to $x_0$, then
the new lists will be disjoint from both solutions and thus the algorithm will
fail.  Apparently, if the algorithm is not to fail at the very beginning
of step (4), it must do some processing to determine \emph{which} list,
of the several satisfying the criterion of step (4), is the ``correct" list
to shrink.

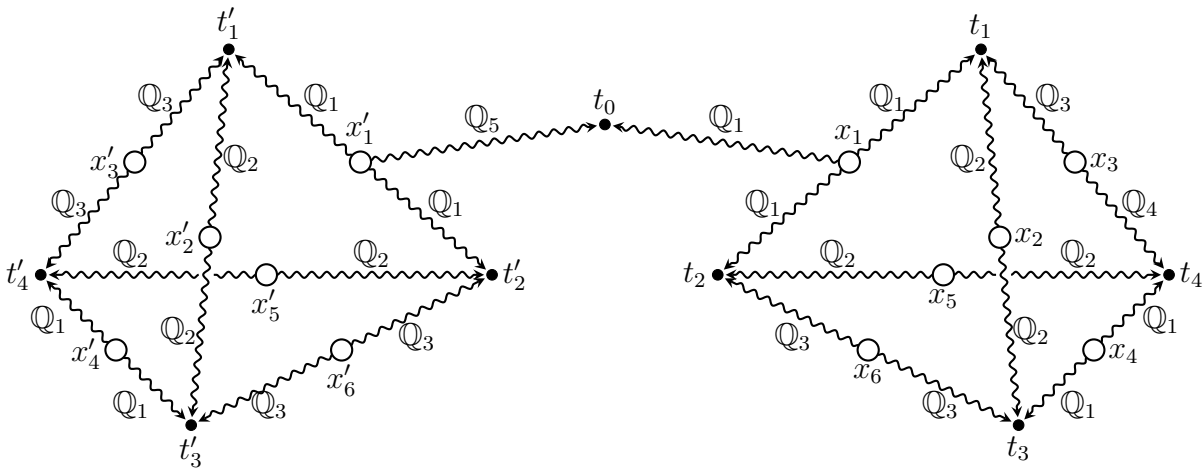
\begin{figure}
\begin{tikzpicture}

\draw [fill] (3.5,3) circle (2.0pt); 
\draw [fill] (0,0) circle (2.0pt);   
\draw [fill] (4,-2) circle (2.0pt);  
\draw [fill] (6,0) circle (2.0pt);   
\draw [fill] (-1.5,2) circle (2.0pt);   

\draw [thick] (1.75,1.5) circle (4.0pt);   
\draw [thick] (3.75,0.5) circle (4.0pt);   
\draw [thick] (2,-1)     circle (4.0pt);   
\draw [thick] (4.75,1.5) circle (4.0pt);   
\draw [thick] (3,0)      circle (4.0pt);   
\draw [thick] (5,-1)     circle (4.0pt);   

\node [anchor=east] at (2.65,2.35) {$\bbQ_1$};
\node [anchor=east] at (3.725,1.55) {$\bbQ_2$};
\node [anchor=west] at (4.075,2.35) {$\bbQ_3$};

\node [anchor=south] at (1.6,-.05) {$\bbQ_2$};
\node [anchor=east] at (1.0,.95) {$\bbQ_1$};
\node [anchor=north] at (1,-.5) {$\bbQ_3$};

\node [anchor=north] at (2.96,-1.4) {$\bbQ_3$};
\node [anchor=north] at (4.8,-1.4) {$\bbQ_1$};
\node [anchor=west] at (3.775,-.75) {$\bbQ_2$};

\node [anchor=west] at (5.225,.95) {$\bbQ_4$};
\node [anchor=south] at (4.8,-.05) {$\bbQ_2$};
\node [anchor=west] at (5.5,-.6) {$\bbQ_1$};

\node [anchor=south] at (.125,1.75) {$\bbQ_1$};

\node [anchor=south] at (3.5,3) {$\Qtop_1$};
\node [anchor=south] at (-1.5,2) {$\Qtop_0$};
\node [anchor=east] at (0,0) {$\Qtop_2$};
\node [anchor=north] at (4,-2) {$\Qtop_3$};
\node [anchor=west] at (6,0) {$\Qtop_4$};

\node [anchor=south] at (1.75,1.6) {$x_1$};
\node [anchor=west] at (3.80,0.5) {$x_2$};
\node [anchor=north] at (2,-1.05) {$x_6$};
\node [anchor=west] at (4.80,1.5) {$x_3$};
\node [anchor=north] at (3,-.05) {$x_5$};
\node [anchor=west] at (5.05,-1.05) {$x_4$};

\draw [->,line before snake=4pt,shorten <=4pt,snake=snake,segment amplitude=.4mm,segment length=2mm,line after snake=5pt,>=stealth,thick,shorten >=3pt] 
(1.75,1.5) -- (-1.5,2);


\draw [->,line before snake=4pt,shorten <=4pt,snake=snake,segment amplitude=.4mm,segment length=2mm,line after snake=5pt,>=stealth,thick,shorten >=3pt] 
(1.75,1.5) -- (3.5,3);

\draw [->,line before snake=4pt,shorten <=4pt,snake=snake,segment amplitude=.4mm,segment length=2mm,line after snake=5pt,>=stealth,thick,shorten >=3pt] 
(1.75,1.5) -- (0,0);

\draw [->,line before snake=4pt,shorten <=4pt,snake=snake,segment amplitude=.4mm,segment length=2mm,line after snake=5pt,>=stealth,thick,shorten >=3pt] 
(3.75,0.5) -- (3.5,3);

\draw [->,line before snake=4pt,shorten <=4pt,snake=snake,segment amplitude=.4mm,segment length=2mm,line after snake=5pt,>=stealth,thick,shorten >=3pt] 
(3.75,0.5) -- (4,-2);

\draw [->,line before snake=4pt,shorten <=4pt,snake=snake,segment amplitude=.4mm,segment length=2mm,line after snake=5pt,>=stealth,thick,shorten >=3pt] 
(2,-1) -- (0,0);

\draw [->,line before snake=4pt,shorten <=4pt,snake=snake,segment amplitude=.4mm,segment length=2mm,line after snake=5pt,>=stealth,thick,shorten >=3pt] 
(2,-1) -- (4,-2);

\draw [->,line before snake=4pt,shorten <=4pt,snake=snake,segment amplitude=.4mm,segment length=2mm,line after snake=5pt,>=stealth,thick,shorten >=3pt] 
(4.75,1.5) -- (3.5,3);

\draw [->,line before snake=4pt,shorten <=4pt,snake=snake,segment amplitude=.4mm,segment length=2mm,line after snake=5pt,>=stealth,thick,shorten >=3pt] 
(4.75,1.5) -- (6,0);

\draw [->,line before snake=4pt,shorten <=4pt,snake=snake,segment amplitude=.4mm,segment length=2mm,line after snake=5pt,>=stealth,thick,shorten >=3pt] 
(3,0) -- (0,0);

\draw [-,thick,line before snake=4pt,shorten <=4pt,snake=snake,segment amplitude=.4mm,segment length=2mm] 
(3,0) -- (3.7,0);

\draw [->,snake=snake,segment amplitude=.4mm,segment length=2mm,line after snake=5pt,>=stealth,thick,shorten >=3pt] 
(3.9,0) -- (6,0); 

\draw [->,line before snake=4pt,shorten <=4pt,snake=snake,segment amplitude=.4mm,segment length=2mm,line after snake=5pt,>=stealth,thick,shorten >=3pt] 
(5,-1) -- (6,0);

\draw [->,line before snake=4pt,shorten <=4pt,snake=snake,segment amplitude=.4mm,segment length=2mm,line after snake=5pt,>=stealth,thick,shorten >=3pt] 
(5,-1) -- (4,-2);


\draw [fill] (-3-3.5,3) circle (2.0pt); 
\draw [fill] (-3-0,0) circle (2.0pt);   
\draw [fill] (-3-4,-2) circle (2.0pt);  
\draw [fill] (-3-6,0) circle (2.0pt);   

\draw [thick] (-3-1.75,1.5) circle (4.0pt);   
\draw [thick] (-3-3.75,0.5) circle (4.0pt);   
\draw [thick] (-3-2,-1)     circle (4.0pt);   
\draw [thick] (-3-4.75,1.5) circle (4.0pt);   
\draw [thick] (-3-3,0)      circle (4.0pt);   
\draw [thick] (-3-5,-1)     circle (4.0pt);   

\node [anchor=west] at (-3-2.65,2.35) {$\bbQ_1$};
\node [anchor=west] at (-3-3.725,1.55) {$\bbQ_2$};
\node [anchor=east] at (-3-4.075,2.35) {$\bbQ_3$};

\node [anchor=south] at (-3-1.6,-.05) {$\bbQ_2$};
\node [anchor=west] at (-3-1.0,.95) {$\bbQ_1$};
\node [anchor=north] at (-3-1,-.5) {$\bbQ_3$};

\node [anchor=north] at (-3-2.96,-1.4) {$\bbQ_3$};
\node [anchor=north] at (-3-4.8,-1.4) {$\bbQ_1$};
\node [anchor=east] at (-3-3.775,-.75) {$\bbQ_2$};

\node [anchor=east] at (-3-5.225,.95) {$\bbQ_3$};
\node [anchor=south] at (-3-4.8,-.05) {$\bbQ_2$};
\node [anchor=east] at (-3-5.5,-.6) {$\bbQ_1$};

\node [anchor=south] at (-3-.125,1.75) {$\bbQ_5$};

\node [anchor=south] at (-3-3.5,3) {$\Qtop_1'$};
\node [anchor=west] at (-3-0,0) {$\Qtop_2'$};
\node [anchor=north] at (-3-4,-2) {$\Qtop_3'$};
\node [anchor=east] at (-3-6,0) {$\Qtop_4'$};

\node [anchor=south] at (-3-1.75,1.6) {$x_1'$};
\node [anchor=east] at (-3-3.80,0.5) {$x_2'$};
\node [anchor=north] at (-3-2,-1.05) {$x_6'$};
\node [anchor=east] at (-3-4.80,1.5) {$x_3'$};
\node [anchor=north] at (-3-3,-.05) {$x_5'$};
\node [anchor=east] at (-3-5.05,-1.05) {$x_4'$};

\draw [->,line before snake=4pt,shorten <=4pt,snake=snake,segment amplitude=.4mm,segment length=2mm,line after snake=5pt,>=stealth,thick,shorten >=3pt] 
(-3-1.75,1.5) -- (-3+1.5,2);


\draw [->,line before snake=4pt,shorten <=4pt,snake=snake,segment amplitude=.4mm,segment length=2mm,line after snake=5pt,>=stealth,thick,shorten >=3pt] 
(-3-1.75,1.5) -- (-3-3.5,3);

\draw [->,line before snake=4pt,shorten <=4pt,snake=snake,segment amplitude=.4mm,segment length=2mm,line after snake=5pt,>=stealth,thick,shorten >=3pt] 
(-3-1.75,1.5) -- (-3-0,0);

\draw [->,line before snake=4pt,shorten <=4pt,snake=snake,segment amplitude=.4mm,segment length=2mm,line after snake=5pt,>=stealth,thick,shorten >=3pt] 
(-3-3.75,0.5) -- (-3-3.5,3);

\draw [->,line before snake=4pt,shorten <=4pt,snake=snake,segment amplitude=.4mm,segment length=2mm,line after snake=5pt,>=stealth,thick,shorten >=3pt] 
(-3-3.75,0.5) -- (-3-4,-2);

\draw [->,line before snake=4pt,shorten <=4pt,snake=snake,segment amplitude=.4mm,segment length=2mm,line after snake=5pt,>=stealth,thick,shorten >=3pt] 
(-3-2,-1) -- (-3-0,0);

\draw [->,line before snake=4pt,shorten <=4pt,snake=snake,segment amplitude=.4mm,segment length=2mm,line after snake=5pt,>=stealth,thick,shorten >=3pt] 
(-3-2,-1) -- (-3-4,-2);

\draw [->,line before snake=4pt,shorten <=4pt,snake=snake,segment amplitude=.4mm,segment length=2mm,line after snake=5pt,>=stealth,thick,shorten >=3pt] 
(-3-4.75,1.5) -- (-3-3.5,3);

\draw [->,line before snake=4pt,shorten <=4pt,snake=snake,segment amplitude=.4mm,segment length=2mm,line after snake=5pt,>=stealth,thick,shorten >=3pt] 
(-3-4.75,1.5) -- (-3-6,0);

\draw [->,line before snake=4pt,shorten <=4pt,snake=snake,segment amplitude=.4mm,segment length=2mm,line after snake=5pt,>=stealth,thick,shorten >=3pt] 
(-3-3,0) -- (-3-0,0);

\draw [-,thick,line before snake=4pt,shorten <=4pt,snake=snake,segment amplitude=.4mm,segment length=2mm] 
(-3-3,0) -- (-3-3.7,0);

\draw [->,snake=snake,segment amplitude=.4mm,segment length=2mm,line after snake=5pt,>=stealth,thick,shorten >=3pt] 
(-3-3.9,0) -- (-3-6,0); 

\draw [->,line before snake=4pt,shorten <=4pt,snake=snake,segment amplitude=.4mm,segment length=2mm,line after snake=5pt,>=stealth,thick,shorten >=3pt] 
(-3-5,-1) -- (-3-6,0);

\draw [->,line before snake=4pt,shorten <=4pt,snake=snake,segment amplitude=.4mm,segment length=2mm,line after snake=5pt,>=stealth,thick,shorten >=3pt] 
(-3-5,-1) -- (-3-4,-2);

\end{tikzpicture}
\caption{Another $\bbG$} \label{fig4}
\end{figure}

This example can be beefed up a bit, by extending $\bbG$ ``on the left" with another
``pyramid" of gadgets, but this time so that the linear constraints on $\{0,1\}$
are consistent (see Figure~\ref{fig4}).
This new system has 8 solutions, all of which equal 2 on $x_1,\ldots,x_6$ and are in
$\{0,1\}$ on $x_1',\ldots,x_6'$.  At the start of step (4), 
the Feder-Kinne-Rafiey algorithm-idea can safely remove the unique
violation of the Mal'tsev property from the list of any of the variables from
the left-hand pyramid, but cannot remove the unique violation of the
Mal'tsev property from the list of any variable from the
right-hand pyramid.
How is the algorithm to decide which list to reduce?  Such are the challenges
facing any attempt to save the Feder-Kinne-Rafiey algorithm-idea.

\subsection*{Acknowledgment}
I am grateful to Arash Rafiey for generously explaining the ideas in
\cite{rkf}.

\end{document}